\documentclass[a4paper,USenglish,cleveref, autoref, thm-restate]{lipics-v2021}

\pdfoutput=1 
\hideLIPIcs  
\nolinenumbers

\title{Improved guarantees for the a priori TSP} 

\author{Jannis Blauth}{Research Inst.\ for Discrete Mathematics, Hausdorff Center for Math., University of Bonn, Germany}{blauth@dm.uni-bonn.de}{https://orcid.org/
	0000-0001-5181-802X}{}

\author{Meike Neuwohner}{Research Inst.\ for Discrete Mathematics, Hausdorff Center for Math., University of Bonn, Germany}{neuwohner@dm.uni-bonn.de}{https://orcid.org/
	0000-0002-3664-3687}{}

\author{Luise Puhlmann}{Research Inst.\ for Discrete Mathematics, Hausdorff Center for Math., University of Bonn, Germany}{puhlmann@dm.uni-bonn.de}{https://orcid.org/
	0009-0001-0776-4586}{}

\author{Jens Vygen}{Research Inst.\ for Discrete Mathematics, Hausdorff Center for Math., University of Bonn, Germany}{vygen@dm.uni-bonn.de}{}{}

\authorrunning{J. Blauth, M. Neuwohner, L. Puhlmann, and J. Vygen} 

\Copyright{Jannis Blauth, Meike Neuwohner, Luise Puhlmann, and Jens Vygen} 



\category{} 

\relatedversion{} 

\EventEditors{John Q. Open and Joan R. Access}
\EventNoEds{2}
\EventLongTitle{42nd Conference on Very Important Topics (CVIT 2016)}
\EventShortTitle{CVIT 2016}
\EventAcronym{CVIT}
\EventYear{2016}
\EventDate{December 24--27, 2016}
\EventLocation{Little Whinging, United Kingdom}
\EventLogo{}
\SeriesVolume{42}
\ArticleNo{23}
\usepackage[utf8]{inputenc}
\usepackage{amsmath,amsfonts, amssymb, amsthm} 
\usepackage{multirow, esdiff}
\usepackage[short]{optidef}
\usepackage[linesnumbered,ruled,vlined]{algorithm2e}
\usepackage{xcolor}
\usepackage{thm-restate}
\usepackage{tikz}
\usepackage{verbatim}
\usepackage{placeins}
\usepackage{dsfont}
\usepackage{xfrac}

\usetikzlibrary{shapes.geometric}

\usepackage[inline]{enumitem}
\setlist[enumerate,1]{label=(\roman*), leftmargin=2.2em}
\setlist[enumerate,2]{label=(\alph*)}

\newlist{stepsroman}{enumerate}{1}
\setlist[stepsroman]{label=(\roman*), leftmargin=2.2em}

\newlist{stepsarabic}{enumerate}{1}
\setlist[stepsarabic]{rightmargin=0.2em, label=\arabic*., ref=\arabic*}

\definecolor{verydarkred}{rgb}{0.5, 0.0, 0.0}
\definecolor{verydarkgreen}{rgb}{0.0, 0.5, 0.0}
\definecolor{darkgreen}{rgb}{0.0, 0.75, 0.0}
\definecolor{darkblue}{rgb}{0.0, 0.0, 0.75}
\theoremstyle{definition}

\newtheorem{comment_}[definition]{Comment}

\newcommand{\Prob}{\mathbb{P}}
\newcommand{\Exp}{\mathbb{E}}
\newcommand{\MR}{\textnormal{MR}}

\newcommand{\Opt}{\textnormal{OPT}}
\newcommand{\OptTSP}{\textnormal{OPT}_{\textsc{tsp}}}

\renewcommand{\epsilon}{\varepsilon}
\renewcommand{\mod}{\, \mathrm{mod} \,}

\begin{document}
\maketitle

\begin{abstract}
We revisit the \textsc{a priori TSP} (with independent activation)
and prove stronger approximation guarantees than were previously known.
In the \textsc{a priori TSP}, we are given a metric space $(V,c)$ 
and an activation probability $p(v)$ for each customer $v\in V$. 
We ask for a TSP tour $T$ for $V$ that minimizes the expected length 
after cutting $T$ short by skipping the inactive customers. 

All known approximation algorithms select a nonempty subset $S$ of the customers and construct a 
\emph{master route solution}, consisting of a TSP tour for $S$  
and two edges connecting every customer $v\in V\setminus S$ to a nearest customer in $S$. 

We address the following questions.
If we randomly sample the subset $S$, what should be the sampling probabilities?
How much worse than the optimum can the best master route solution be?
The answers to these questions (we provide almost matching lower and upper bounds)
lead to improved approximation guarantees:
less than 3.1 with randomized sampling, and less than 5.9 
with a deterministic polynomial-time algorithm.
\end{abstract}

\section{Introduction}\label{sec:introduction}

Many algorithms for stochastic discrete optimization problems sample a sub-instance, 
solve the resulting deterministic problem (often by some approximation algorithm), 
and extend this solution to the original instance \cite{EGRS10,GarGLS08,GKPR07,GKT03,GPRS04,ShmT08}. 
A nice and well-studied example is the \textsc{a priori Traveling Salesperson Problem} (\textsc{a priori TSP}), which is the focus of this paper.
What guarantee can we obtain by such an approach, even if we take an optimal sample? 
If we sample randomly, according to which distribution?
What guarantee can we obtain by a deterministic polynomial-time algorithm?
These are the questions addressed in this paper.

In the \textsc{a priori TSP} (with independent activation), 
we are given a (semi-)metric space $(V,c)$; the elements of $V$ are called \emph{customers}.
Each customer $v$ comes with an \emph{activation probability} $0<p(v)\le 1$, so it will be active independently with probablity $p(v)$.
However, we need to design a TSP tour $T$ (visiting all of $V$) \emph{before} knowing which customers will be active.
\emph{After} we know which customers are active we can cut the tour $T$ short by skipping the inactive customers. 
The goal is to minimize the expected cost of the resulting tour (visiting the active customers). 

Note that computing an optimum a priori tour is APX-hard as the metric TSP is APX-hard~\cite{APXhardness}, which is the special case where all activation probabilities are $1$. We study approximation algorithms. A \emph{$\rho$-approximation algorithm} for the \textsc{a priori TSP} is a polynomial-time algorithm that computes a tour of expected cost at most $\rho \cdot \Opt$ for any given instance, where $\Opt$ denotes the expected cost of an optimum a priori tour.

Shmoys and Talwar~\cite{ShmT08} devised a randomized 4-approximation algorithm 
and a deterministic 8-approximation algorithm. 
A randomized constant-factor approximation algorithm was discovered independently by 
Garg, Gupta, Leonardi and Sankowski \cite{GarGLS08}.
The randomized Shmoys--Talwar algorithm easily improves to a 3.5-approximation by using the
Christofides--Serdyukov algorithm instead of the double tree algorithm as a subroutine for TSP
(as noted by \cite{EeS18}), and slightly better using the new Karlin--Klein--Oveis Gharan algorithm \cite{KarKO21}.
The deterministic algorithm was improved to a 6.5-approximation by van Zuylen~\cite{Zuy11};
a slight improvement of this guarantee follows from the recent deterministic version of the
Karlin--Klein--Oveis Gharan algorithm \cite{KarKO23}.

\begin{figure}[ht]
\begin{center}
\begin{tikzpicture}[scale=0.85, thick]
\tikzstyle{mvertex}=[darkgreen,circle,draw,minimum size=6,inner sep=0pt]
\tikzstyle{avertex}=[blue,circle,fill,minimum size=6,inner sep=0pt]
\tikzstyle{ivertex}=[blue,circle,draw,minimum size=6,inner sep=0pt]
\tikzstyle{bedge}=[darkgreen,line width=2]
\tikzstyle{redge}=[red,line width=1]
\begin{scope}[shift={(0,0)}]
\node[mvertex] (s)  at (0.1, 1) {};
\node[mvertex] (m1)  at (1, 3) {};
\node[mvertex] (m2)  at (3, 2.5) {};
\node[mvertex] (m3)  at (2.9, 1) {};
\node[mvertex] (m4)  at (1.5, 1.5) {};
\draw[bedge] (s) -- (m1);
\draw[bedge] (m1) -- (m2);
\draw[bedge] (m2) -- (m3);
\draw[bedge] (m3) -- (m4);
\draw[bedge] (m4) -- (s);
\node[ivertex] (i01)  at (-0.9, 1.5) {};
\node[ivertex] (i11)  at (-0.4, 2.9) {};
\node[ivertex] (i21)  at (4.2, 2.9) {};
\node[ivertex] (i31)  at (4.7, 1.8) {};
\node[ivertex] (i32)  at (4.6, 0.2) {};
\node[ivertex] (i33)  at (2.5, 0) {};
\draw[redge] (i01) to[bend left=20] (s);
\draw[redge] (s) to[bend left=20] (i01);
\draw[redge] (i11) to[bend left=15] (m1);
\draw[redge] (m1) to[bend left=15] (i11);
\draw[redge] (i21) to[bend left=15] (m2);
\draw[redge] (m2) to[bend left=15] (i21);
\draw[redge] (i31) to[bend left=10] (m3);
\draw[redge] (m3) to[bend left=10] (i31);
\draw[redge] (i32) to[bend left=10] (m3);
\draw[redge] (m3) to[bend left=10] (i32);
\draw[redge] (i33) to[bend left=20] (m3);
\draw[redge] (m3) to[bend left=20] (i33);
\end{scope}

\begin{scope}[shift={(9,0)}]
\node[avertex] (s)  at (0.1, 1) {};
\node[avertex] (m1)  at (1, 3) {};
\node[ivertex] (m2)  at (3, 2.5) {};
\node[ivertex] (m3)  at (2.9, 1) {};
\node[ivertex] (m4)  at (1.5, 1.5) {};
\draw[bedge] (s) -- (m1);
\draw[bedge] (m1) -- (m2);
\draw[bedge] (m2) -- (m3);
\draw[bedge] (m3) -- (m4);
\draw[bedge] (m4) -- (s);
\node[ivertex] (i01)  at (-0.9, 1.5) {};
\node[avertex] (i11)  at (-0.4, 2.9) {};
\node[ivertex] (i21)  at (4.2, 2.9) {};
\node[avertex] (i31)  at (4.7, 1.8) {};
\node[ivertex] (i32)  at (4.6, 0.2) {};
\node[avertex] (i33)  at (2.5, 0) {};
\draw[redge] (i11) to[bend left=15] (m1);
\draw[redge] (m1) to[bend left=15] (i11);
\draw[redge] (i31) to[bend left=10] (m3);
\draw[redge] (m3) to[bend left=10] (i31);
\draw[redge] (i33) to[bend left=20] (m3);
\draw[redge] (m3) to[bend left=20] (i33);
\end{scope}

\end{tikzpicture}
\end{center}
\caption{
Left: A master route solution with a master tour (green, thick) and connections of the other customers to that master tour (red, curved). 
Right: After knowing which customers are active (filled), 
the master route solution reduces to a tour visiting all of the master tour and the other active customers.
\label{fig:initial_example}}
\end{figure}
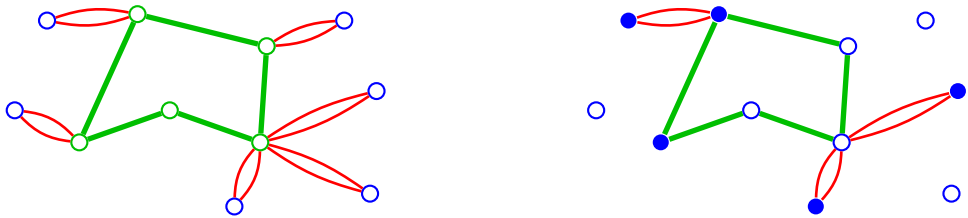

All known approximation algorithms for the \textsc{a priori TSP} are of the following type.
Select a nonempty subset $S$ of customers and find a TSP tour for $S$ (the \emph{master tour}).
Connect each other customer $v\in V\setminus S$ with a pair of parallel edges
to a nearest point $\mu(v)$ in the master tour. We call this a \emph{master route solution}.
Once we know the set of active customers, we pay for the entire master tour 
(pretending to visit also its inactive customers!)\ 
and pay $2c(\mu(v),v)$ for each active customer $v$ outside $S$ 
to cover the round trip visiting $v$ from $\mu(v)$. 
See \cref{fig:initial_example} for an example.
Of course, we could cut the resulting tour shorter (we visit some inactive customers, and 
we visit some customers several times), but we will not account for this possible gain
(unless fewer than two customers are active).

\subsection{Motivating questions} \label{sec:overview}

We start by reviewing the randomized algorithm by Shmoys and Talwar~\cite{ShmT08}.
If fewer than two customers are active, any a priori tour can be cut short to a single point, resulting in cost zero.
The algorithm by Shmoys and Talwar~\cite{ShmT08} 
selects each customer $v$ independently into $S$ with probability $p(v)$: exactly the activation probability. 
Assuming that the resulting set $S$ is nonempty, there exists an associated master route solution with expected cost at most
\begin{equation*}
\MR(S) \ \coloneqq \ \ \Exp_{A\sim p} \left[ \mathds{1}_{|A|\ge 2} \cdot \left( \OptTSP(S,c)+2\cdot\sum_{v\in A}c(v,S) \right) \right] .
\end{equation*}
Here $\OptTSP(S,c)$ denotes the length of an optimum TSP tour for $S$, 
and $c(v,S)=\min\{c(v,s): s\in S\}$ denotes the distance between $v$ and a nearest customer in $S$ (which is zero if $v\in S$); 
moreover, $\Exp_{A\sim p}$ denotes the expectation
when the set $A$ of active customers is sampled with respect to the given activation probabilities. Later on, $\Prob_{A\sim p}$ is used analogously.
We multiply with $ \mathds{1}_{|A|\ge 2}$ (which is 1 if $|A|\ge 2$ and 0 otherwise)
because the cost of the solution is zero if fewer than two customers are active.

If there is a customer $d$ with $p(d)=1$ (a \emph{depot}), then $S$ is never empty and we can bound
\begin{align*}
\MR(S) &\ \le \  \OptTSP(S,c)+2\cdot\Exp_{A\sim p}\left[\sum_{v\in A\setminus\{d\}}c(v,S\setminus\{v\})\right] .
\end{align*}

Note that the above upper bound also accounts for connecting active customers in $S$ to the nearest other customer in $S$, 
which is not necessary but will allow the following. Taking the expectation over the random choice of $S$,
an upper bound on the expected cost of that master route solution is
\vspace*{-2mm}
\begin{equation*}
\Exp_{S\sim p}\left[\MR(S)\right] \ \le \ 
\Exp_{S\sim p} \left[\OptTSP(S,c) \right] + 2 \cdot \Exp_{S\sim p} \left[ \sum_{v\in S\setminus\{d\}} c(v,S\setminus\{v\}) \right]
\end{equation*}
as the probability distributions to choose $S$ and $A$ are identical and the vertices are sampled independently. 
Since $\sum_{v\in S\setminus\{d\}} c(v,S\setminus\{v\}) \le \OptTSP(S,c)$ for all $S$ and $\Exp_{S\sim p} \left[ \OptTSP(S,c) \right] \le \Opt$, 
where $\Opt$ again denotes the expected cost of an optimum a priori tour, this yields 
\begin{equation}\label{eq:upperbound3}
\Exp_{S\sim p} \left[ \MR(S) \right] \ \le \ 3\cdot\Opt.
\end{equation}

The work of Shmoys and Talwar \cite{ShmT08} implies that \eqref{eq:upperbound3} also holds when there is no depot
and when we take the conditional expectation under the condition that $|S|\ge 2$ (see also \cite{Zuy11}).
The Shmoys--Talwar algorithm cannot find an optimum TSP tour for $S$ 
but uses the double tree algorithm with approximation guarantee 2. 
As noted by \cite{EeS18}, one can as well use the Christofides--Serdyukov algorithm 
with approximation guarantee $\frac{3}{2}$, or in fact
any $\alpha$-approximation algorithm for TSP. 
Then the expected cost of the resulting master route solution is at most
$(\alpha+2)\cdot\Opt$. 

This motivates the following questions:
\begin{enumerate}
\item Is it optimal to sample $S$ with exactly the activation probabilities (which is crucially used in the above analysis), 
or can we improve on the factor $\alpha+2$ by sampling fewer or more?
\item How bad can the best master route solution be? 
We will call this the \emph{master route ratio}: by the Shmoys--Talwar analysis, it is at most 3.
\item Can we obtain an approximation guarantee equal to the master route ratio by a master route solution based on random sampling,
assuming that we can find optimum TSP tours? 
What is the best we can achieve with a $\frac{3}{2}$-approximation algorithm for TSP?
\item Can we obtain a better deterministic algorithm without a better TSP algorithm?
\end{enumerate}
We give almost complete answers to all these questions.

\subsection{Our results}
\label{sec:our_results}

The possibility that we sample the empty set or that no customer is active causes significant complications. The previous works \cite{ShmT08} and \cite{Zuy11} gave ad hoc proofs that \emph{their} 
algorithms (which are also formulated with a depot) generalize to the non-depot case. We aim for a general reduction, losing only an arbitrarily small constant:
Fortunately, instances in which the expected number of active customers is small can be solved easily with
an approximation factor $3+\epsilon$ (for any $\epsilon>0$; similar to \cite{EGRS10}; cf.\ \cref{lemma_low_activity}), 
and hence much better than the known guarantees.
For instances with a large expected number of active customers, one can assume without loss of generality
(with an arbitrarily small loss) that there is a customer $d$ that is always active, 
i.e., $p(d)=1$ (\cref{lemma:depot_when_highactivity}). 
So we assume this henceforth and call $d$ the depot. We summarize (and refer to \cref{section:wlog_depot} for the proof):

\begin{theorem} \label{thm:can_assume_depot}
	Let $\epsilon > 0$ and $\rho\ge 3$ be constants.
	If there exists a (randomized) polynomial-time $\rho$-approximation algorithm 
	for instances $(V,c,p)$ of the \textsc{a priori TSP} that have a depot (i.e., a customer $d$ with $p(d) = 1$), 
	then there is a (randomized) polynomial-time $(\rho+\epsilon)$-approximation algorithm 
	for general instances of the \textsc{a priori TSP}.
\end{theorem}

The Shmoys-Talwar algorithm \cite{ShmT08} 
includes a customer $v$ into $S$ with probability $p(v)$: 
the sampling probability is exactly the activation probability. 
Although this is natural and allows for the simple analysis in \cref{sec:overview} (assuming a depot), we show that this is not optimal. 
Decreasing the probability of including a customer into the master tour improves the approximation guarantee. 
To be more precise, in \cref{sec:sampling_ub} and \cref{sec:sampling_lb}, 
we analyze the following \emph{sampling algorithm} for \textsc{a priori TSP} instances with depot. 
Let $f \colon (0,1]\to [0,1]$ with $f(1)=1$. 
\begin{enumerate}
	\item Sample a subset $S\subseteq V$ by including every customer $v$ independently with probability $f(p(v))$.
	\item Call an $\alpha$-approximation algorithm for (metric) TSP in order to compute a TSP tour for $S$, which serves as master tour.
	\item Connect every customer outside $S$ to the nearest customer in $S$ by a pair of parallel edges. 
\end{enumerate}
For a given instance this algorithm has expected approximation ratio at most
\begin{equation} \label{eq:apxratio_of_sampling}
\frac{1}{\Opt} \cdot \Exp_{S\sim f\circ p} \left[ \alpha \cdot \OptTSP(S,c) + 2 \cdot \sum_{v\in V} p(v) \cdot c(v,S) \right]
\end{equation}
(where $\frac{0}{0}\coloneqq 1$).
Shmoys and Talwar \cite{ShmT08} used the identity function $f(p)=p$. 
It is easy to construct examples where sampling less or more is better.
For example, if $c(v,w)=1$ for all $v,w\in V$ with $v\not=w$ (and all activation probabilities except for the depot are tiny),
it is best to include only the depot in the master tour: 
this yields an approximation ratio of 2 instead of 3.
On the other hand, if $V=\{v_0,\ldots,v_{n-1}\}$ and $c(v_i,v_j)=\min\{j-i,n+i-j\}$ for $i<j$
(i.e., $(V,c)$ is the metric closure of a cycle), the more we sample, the better.
However, even if we choose $f$ depending on the instance, there is a limit on what we can achieve:

\begin{theorem} \label{thm:sampling_lb}
	No matter how $f$ is chosen, even depending on the instance in an arbitrary way, 
	the sampling algorithm has no better approximation ratio than 
	\begin{itemize}
	\item	$2.655$ even if it computes an optimum TSP tour on the sampled customers;
	\item $3.049$ assuming that we never compute a TSP tour on the sampled customers of cost less than $1.4999$ times the cost of an optimum tour. 
	\end{itemize}
\end{theorem}

See \cref{sec:sampling_lb} for the proof.
We do not have a matching upper bound, but we come close. For $\alpha=1.5$ we prove (in \cref{sec:sampling_ub}):

\begin{theorem}\label{thm:ub_sampling}
	For $\alpha = 1.5$ and $f(p)=1-(1-p)^{\sigma}$ with $\sigma = 0.663$, 
	the sampling algorithm for \textsc{a priori TSP} instances with depot has approximation guarantee less than 3.1.
\end{theorem}	
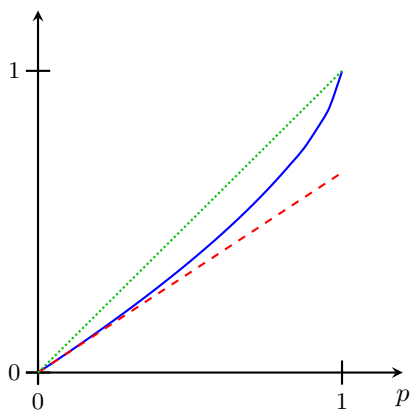
\begin{figure}
\begin{tikzpicture}[scale=0.8]
	\draw[-stealth, thick] (-0.2, 0) to node[below=1mm, pos =1]{$p$} (6,0);
	\draw[-stealth, thick] (0,-0.2) to (0,6);
\draw[thick] (0, -0.2) to node[below=1.5mm]{\small $0$} (0, 0.2);
\draw[thick] (-0.2, 0) to node[left=1mm]{\small $0$} (0.2, 0);
\draw[thick] (5, -0.2) to node[below=1.5mm]{\small $1$} (5, 0.2);
\draw[thick] (-0.2, 5) to node[left=1mm]{\small $1$} (0.2, 5);
\draw[domain=0:5, thick, blue, smooth] plot ({\x},{5*(1-(1-0.2*\x)^0.663)});
\draw[domain=0:5, thick, red, dashed] plot ({\x},{0.663*\x});
\draw[domain=0:5, thick, darkgreen, densely dotted] plot ({\x},{\x});
\end{tikzpicture}
\caption{The function $p\mapsto 1-(1-p)^\sigma$ with $\sigma=0.663$ (blue, solid) defines the sampling probability in \cref{thm:ub_sampling}, which is always at most the identity function (green, dotted), and for small $p$ approximately equal to $p\mapsto \sigma\cdot p$ (red, dashed).\label{figure:f_sampling}}
\end{figure}
\cref{figure:f_sampling} shows this function $f$. Together with \cref{thm:can_assume_depot} this immediately implies one of our main results:

\begin{corollary}
There is a randomized $3.1$-approximation algorithm for \textsc{a priori TSP}. \hfill\qed
\end{corollary}

We conjecture that the bounds in \cref{thm:sampling_lb} are actually attained by the sampling algorithm 
with $f(p)=1-(1-p)^{\sigma}$, independent of the instance,
where $\sigma$ is a positive constant that depends on $\alpha$ only. 
See \cref{comment:conjecture_sampling} for details.

Having explored the limits of the random sampling approach, 
one might ask what is the limit of choosing an \emph{optimal} master route solution.
By van Zuylen's work \cite{Zuy11},
the answer to this question is the key to obtain a better deterministic approximation algorithm 
(see \cref{section:deterministic_via_masterrouteratio}). 
Let us define:

\begin{definition}[master route ratio]
The \emph{master route ratio} is defined to be the supremum of
$$\frac{\min  \left\{ \MR(S) : \emptyset\not= S\subseteq V\right\}}{\Opt}$$
taken over all \textsc{a priori TSP} instances (where $\frac{0}{0}\coloneqq1$).
\end{definition}

It is very easy to see that the master route ratio is at least 2 (for example, if $c(v,w)=1$ for all $v,w\in V$ with $v\not=w$).
By the Shmoys--Talwar analysis, it is at most 3. 
We will show in \cref{sec:mastrerouteratio_ub} (upper bound) and \cref{subsection:lowerboundonmasterrouteratio} (lower bound):

\begin{theorem}\label{thm:mrr}
	The master route ratio for \textsc{a priori TSP} instances with depot
	is at least $\frac{1}{1-e^{-1/2}}>2.541$ and less than $2.6$. 
\end{theorem}	
We conjecture that the master route ratio is exactly $\frac{1}{1-e^{-1/2}}$.

As van Zuylen's \cite{Zuy11} analysis reveals 
(cf.\ \cref{thm:masterrouteratio_implies_deterministic} in \cref{section:deterministic_via_masterrouteratio}),
her algorithm is a $(2+\alpha\rho)$-approximation algorithm
if the master route ratio is $\rho$ and we have an algorithm for TSP that guarantees to produce
a tour of cost at most $\alpha$ times the value of the subtour relaxation.
So our new upper bound on the master route ratio immediately implies a better guarantee
(combining Theorems~\ref{thm:can_assume_depot}, \ref{thm:mrr}, and \ref{thm:masterrouteratio_implies_deterministic}
with $\alpha = \frac{3}{2}$ \cite{Wol80}):

\begin{corollary} \label{cor:approximation_ratio_deterministic}
	There is a deterministic 5.9-approximation algorithm for \textsc{a priori TSP}.
	\hfill\qed
\end{corollary}
\vspace*{-8mm}

\subsection{Our techniques}\label{sec:our_techniques}

The lower bounds (\cref{thm:sampling_lb} and the lower bound in \cref{thm:mrr}) are obtained by analyzing simple examples. 
The main technical difficulty is in proving the upper bounds.

To prove \cref{thm:ub_sampling} and the upper bound in \cref{thm:mrr}, we will show that it suffices to consider instances 
in which all customers (except the depot) have the same tiny activation probability. 
We call these instances \emph{normalized}.

\begin{definition}
	Let $\varepsilon>0$.
	An instance $(V,c,p)$ of \textsc{a priori TSP} 
	is called \emph{$\varepsilon$-normalized} if the instance contains a depot $d\in V$ (with $p(d)=1$), 
	and $p(v)=\varepsilon$ for all $v\in V\setminus\{d\}$.
\end{definition}

Given an instance of  \textsc{a priori TSP} with a depot $d$, one can transform it to a normalized instance 
by replacing each customer $v \in V \setminus \{d\}$ by many copies, each with the same tiny activation probability, 
such that the probability that at least one of these copies is active is roughly $p(v)$. 
This way, the master route ratio and the approximation guarantee of the sampling algorithm can only get worse. 
More precisely, we show (in \cref{section:reduce_to_normalized}):

\begin{lemma} \label{lemma:wlog_normalized}
	Let $(\varepsilon_i)_{i \in \mathbb{N}} \in (0,1]^{\mathbb{N}}$ with $\lim_{i \rightarrow \infty} \varepsilon_i = 0$. Let $\mathcal{I}$ be the class of all $\varepsilon$-normalized instances with $\varepsilon = \varepsilon_i$ for some $i \in \mathbb{N}$. Then  
	\begin{enumerate}
		\item \label{lemma:wlog_normalized_mr}The master route ratio is the same when restricting it to instances in $\mathcal{I}$
		and when restricting it to all instances with depot.
		\item \label{lemma:wlog_normalized_sampling} Let $\sigma \in (0,1)$. Every upper bound on \eqref{eq:apxratio_of_sampling} 
		for $f(p) = \sigma p \enskip\forall p\in (0,1)$ for all instances in $\mathcal{I}$
		implies the same upper bound on \eqref{eq:apxratio_of_sampling} 
		for $f(p) = 1-(1-p)^\sigma$ for arbitrary instances with depot.
	\end{enumerate}
\end{lemma}	

On a high level, our proofs of \cref{thm:ub_sampling} and \cref{thm:mrr} are similar. 
In both cases we will design a linear program that encodes the metric $c$ by variables and minimizes the 
expected cost of an optimum a priori tour
subject to (a relaxation of) the constraint that the expected cost of the output of the sampling algorithm is at least 1 (for \cref{thm:ub_sampling}) or 
the expected cost of any master route solution is at least 1 (for \cref{thm:mrr}), respectively.
Then the reciprocals of the LP values yield the desired upper bounds.

However, this approach has to overcome several obstacles.
First, it is not obvious how to encode the metric $c$ by finitely many variables, given that we need to consider arbitrary instance sizes.
We do this by fixing an optimum a priori tour $T^*$ (a cyclic order of the customers) 
and carefully aggregating distances of customer pairs with the same number of hops in between on $T^*$.
Of course we exploit the structure of normalized instances.

In the end, we will (almost) ignore variables that correspond to a very large number of hops 
(where it is very unlikely that none of the customers ``in between'' is active). 
These variables have negligible impact because the probability that these edges occur decreases exponentially 
with increasing number of hops on $T^*$, whereas the average length of these edges can only grow linearly due to the triangle inequality. 

The next idea is to consider certain structured solutions only. 
Rather than connecting a customer $v$ that is not in the master tour to the \emph{nearest} customer $\mu(v)$ in the master tour, 
we consider only two possible members of the master tour:
we traverse $T^*$ from $v$ in each of the two possible directions, and consider  
the first customer that we meet and that is contained in our master tour. 
None of these two may be a nearest one in the master tour, but we still obtain an upper bound.
For bounding the master route ratio, we will in addition only consider master tours whose customers are equidistantly distributed on $T^*$ (except for the depot).

In this way, we obtain an optimization problem for a fixed uniform activation probability $p$ (i.e., for $p$-normalized instances). 
However, we must let $p\to 0$ according to \cref{lemma:wlog_normalized} and hence need a description that is independent of $p$. 
This is another major obstacle. To overcome it, we use a second level of aggregation (buckets, rounding the number of hops to integer multiples of, say, $\frac{1}{100p}$).
However, this causes several difficulties.
In the case of the sampling algorithm, describing the expected cost of the output of the sampling algorithm in terms of the buckets is nontrivial.
In case of the master route ratio, the same holds for master route solutions and actually requires a third level of aggregation (bucket intervals).

In the end, we obtain (in both cases) a single, relatively compact, linear program 
that yields an upper bound for all instance sizes and all activation probabilities from a sequence that converges to zero. 
We solve the dual LP numerically and just need to check feasibility to prove the desired upper bounds.

\subsection{Further related work}

The TSP has also been studied under the aspect of robust optimization, where the set of customers that need to be visited is known in advance, 
but the edge lengths are chosen probabilistically or even adversarially~\cite{GaneshMaggsPanigrahi,ToHaPo14}. 
The a priori optimization problem where the set of customers is chosen adversarially is known as universal TSP~\cite{GarGLS08,LowerBoundsForUniversalAndAPriori,SchSh08}. 
The probabilistic version that we consider was introduced by Jaillet~\cite{Jai85} and Bertsimas~\cite{Ber88}. 
Since then, various aspects of the problem have been investigated, including the asymptotic behavior of random instances~\cite{Ber88,BerJaiA90,BowFB03,Jai85,Jai88}, online variants~\cite{GarGLS08}, or exact algorithms~\cite{AbKhaBell17}. 
Approximation algorithms have also been studied for general probability distributions~\cite{VANEE2018331,LowerBoundsForUniversalAndAPriori,SchSh08}.

Other problems that have been considered in an a priori setting include vehicle routing, traveling repairman, 
Steiner tree, and  
network design~\cite{EeS18,EGRS10,FerStei20,GarGLS08,GKPR07,GKT03,GPRS04, NGN20}. 
However, none of these works managed to determine the approximation guarantee of their algorithms exactly.

Previous approaches to design a linear program that yields the approximation ratio of a certain algorithm for some optimization problem (e.g., ~\cite{Goemans1998,DualFitting})
typically required an infinite family of linear programs 
and could not obtain a bound for general instances by just solving a single linear program.

\section{Upper bound on the approximation ratio of random sampling}
\label{sec:sampling_ub}

In this section we will prove \cref{thm:ub_sampling}. 
As mentioned earlier, we will design a single linear program such that the reciprocal of its optimum value is an upper bound on the approximation ratio of the sampling algorithm for a certain class of normalized instances. 
For this sake, let $\beta, b_0 > 0$ be constants that we will choose later.
We will consider $\varepsilon$-normalized instances where $\varepsilon$ is of the form $\varepsilon = \frac{\beta}{b}$ for some odd integer $b \ge b_0$. The meaning of these constants will become clear in \cref{section:sampling_singleLP}.
For such instances we will obtain an upper bound on the approximation ratio of the sampling algorithm, 
when sampling each customer with probability $\sigma p$ for $\sigma= 0.663$ (in addition to the depot). 
Combined with \cref{lemma:wlog_normalized}, this immediately yields the same upper bound on the approximation guarantee 
of the sampling algorithm that samples each customer $v$ with probability $1-(1-p(v))^\sigma$
for arbitrary \textsc{a priori TSP} instances with depot.

\subsection{An optimization problem to bound the approximation ratio}\label{sec:sampling_OP}

In this section, we first describe an upper bound for all $p$-normalized instances (for a fixed uniform activation probability $p$)
by a single optimization problem.
We will consider the algorithm that samples each customer with probability $\sigma p$.
Let $T^*$ be a fixed optimum a priori tour, with customers appearing in the order $v_0,v_1,\ldots,v_{n-1}$; 
here $v_0$ denotes the depot. Let $v_i\coloneqq v_0$ for $i<0$ or $i>n-1$.
For $k\in\mathbb{Z}_{\ge 1}$ we define 
\[
C_k \ \coloneqq \ p^2 \cdot \sum_{j\in\mathbb{Z}} c(v_j,v_{j+k}).
\]

\begin{figure}[h]
	\begin{center}
		\begin{tikzpicture}
			\node[draw,minimum size=4cm,regular polygon,regular polygon sides=9, thick] (a) {};
			\draw[red!80!black, ultra thick] (a.corner 1) to[bend right] (a.corner 2)  -- (a.corner 5) -- (a.corner 8) -- (a.corner 1)
			-- (a.corner 3) -- (a.corner 6) -- (a.corner 9) to[bend right] (a.corner 1)
			-- (a.corner 4) -- (a.corner 7) -- (a.corner 1);
			\draw (a.corner 9) -- (a.corner 1) -- (a.corner 2);
			\fill[white] (a.corner 1) circle[radius=3pt];
			\draw (a.corner 1) circle[radius=3pt];
			\foreach \x in {2,3,...,9}
			\fill (a.corner \x) circle[radius=3pt];
		\end{tikzpicture}
	\end{center}
	\caption{The depot $v_0$ is the white circle at the top; the tour $T^*$ is drawn in black. 
	Adding up the costs of the edges marked in red gives $\frac{C_3}{p^2}$. \label{fig:C3}}
\end{figure}
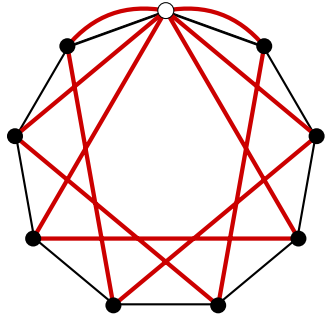

Observe that only finitely many summands are nonzero. See Figure~\ref{fig:C3} for an example. 
Since $c$ is a metric, the numbers $C_k$ are nonnegative and satisfy the triangle inequality, that is, for all $i,j \ge 1$
	\begin{equation} \label{eq:triangleineqsampling}
		C_{i+j} \ \leq \ C_i+C_j .
	\end{equation}

Moreover, we can express the expected cost of $T^*$ in terms of the $C_i$.
\begin{proposition}\label{prop:opt_C_i}
The expected cost of $T^*$ is exactly
\begin{equation} \label{eq:costofoptimumsolutionsampling}
	\sum_{i=1}^{\infty} (1-p)^{i-1}\cdot C_i .
\end{equation}
\end{proposition}
\begin{proof}
	Let $1 \le i \le n-2$ and $1 \le j \le n-i-1$.
	Then $v_j$ and $v_{j+i}$ are consecutive active customers with probability $p^2\cdot (1-p)^{i-1}$; 
	note that the cost of the edge $\{v_j,v_{j+i}\}$ is counted with exactly the same coefficient in \eqref{eq:costofoptimumsolutionsampling}. 
	Moreover, for $1 \le j \le n-1$, $v_j$ is the first active customer after the depot with probability $p\cdot (1-p)^{j-1}$,
	and the cost of the edge $\{v_0,v_j\}$ is counted 
	$\sum_{i=j}^{\infty} p^2\cdot (1-p)^{i-1} = p\cdot (1-p)^{j-1}$ times in  \eqref{eq:costofoptimumsolutionsampling}.
	By symmetry, the terms also match for the last active customer before the depot.
\end{proof}

We now consider the master route solution resulting from sampling each customer with probability $\sigma p$ (in addition to the depot). 
Let $\alpha$ again denote the approximation guarantee of the TSP algorithm that we use. 
We will now show that the expected cost of this master route solution is at most
\begin{equation} \label{eq:costofsampledmasterroutesolutionsampling}
	\sigma^2 \sum_{k=1}^\infty (1-\sigma p)^{k-1}\cdot \left( \alpha \cdot C_k + 2p \cdot \sum_{i=1}^{k-1} \min\left\{\big. C_i,\, C_{k-i}\right\} \right)
	.
\end{equation}

By the same argumentation as in the proof of \cref{prop:opt_C_i}, the master tour has expected cost at most
\begin{align*}
	\alpha\cdot\Exp_{S\sim q}[c(T^*[S])] \ = \ \alpha\cdot \sigma^2\cdot\sum_{k=1}^\infty (1-\sigma p)^{k-1}\cdot C_k ,
\end{align*}
where $q(v)=\sigma p$ for all $v\in V\setminus \{d\}$ and $q(d)=1$.

Next we bound the expected cost of connecting the active customers to the master tour. 
Instead of connecting $v$ to the nearest customer in the master tour, we consider only two options:
the first sampled customer that we meet when traversing $T^*$ from $v$ in either direction.
Note that sampling $v_0$ with probability $1$ is equivalent to sampling each $v_j$ with $j\leq 0$ and $j\geq n$ with probability $\sigma p$. 
Now, for $j\in\mathbb{Z}$ and $k\geq 2$, the probability that $v_j$ and $v_{j+k}$ are sampled, but none of the intermediate customers is, 
equals $(\sigma p)^2\cdot (1-\sigma p)^{k-1}$. 
In this case, the total expected cost of connecting the intermediate active customers 
can be bounded by $2 p\cdot \sum_{i=1}^{k-1}\min\{c(v_j,v_{j+i}),c(v_{j+i},v_{j+k})\}$.
Thus we can bound the expected cost of connecting all active customers to the master tour by
\begin{align*}
	&\sum_{k=2}^\infty \sigma^2 p^2 \cdot (1-\sigma p)^{k-1} \cdot\sum_{j\in\mathbb{Z}}  2 p\cdot\sum_{i=1}^{k-1} \cdot \min\left\{\big. c(v_j,v_{j+i}),\, c(v_{j+i},v_{j+k}) \right\} \\
	&\ \le \ 2 \sigma^2 p^3 \cdot \sum_{k=1}^\infty(1-\sigma p)^{k-1} \cdot \sum_{i=1}^{k-1} \min\left\{\sum_{j\in\mathbb{Z}} c(v_j,v_{j+i}),\, \sum_{j\in\mathbb{Z}} c(v_{j+i},v_{j+k}) \right\} \\
	&\ = \  2 \sigma^2 p^3 \cdot \sum_{k=1}^\infty(1-\sigma p)^{k-1} \cdot \sum_{i=1}^{k-1} \min\left\{ \sum_{j\in\mathbb{Z}} c(v_j,v_{j+i}),\, \sum_{j\in\mathbb{Z}} c(v_{j},v_{j+(k-i)}) \right\} \\
	&\ = \ 2 \sigma^2 p \cdot \sum_{k=1}^\infty(1-\sigma p)^{k-1} \cdot \sum_{i=1}^{k-1}\min\left\{\big. C_i,\, C_{k-i}\right\}.
\end{align*}

We conclude that the ratio of \eqref{eq:costofsampledmasterroutesolutionsampling} to \eqref{eq:costofoptimumsolutionsampling} 
is an upper bound on the approximation guarantee of the sampling algorithm for that instance. 
Note that the number of customers appears neither in \eqref{eq:costofoptimumsolutionsampling} nor in \eqref{eq:costofsampledmasterroutesolutionsampling}.
In other words, minimizing \eqref{eq:costofoptimumsolutionsampling} subject to the constraints that
\eqref{eq:costofsampledmasterroutesolutionsampling} is equal to 1 and the $C_i$ are nonnegative and
satisfy the triangle inequality \eqref{eq:triangleineqsampling} yields the reciprocal of an upper bound
on the approximation guarantee of the sampling algorithm on all $p\,$-normalized instances. 
We arrive at the following optimization problem:
\begin{align}
	\min \sum_{i=1}^{\infty} (1-p)^{i-1}\cdot C_i  &&  \tag{Sampling-OP} \label{eq:exact_lp_for_samplingalgo} \\
	\text{subject to } \hspace*{5cm} 
	C_i &\ \geq \ 0 &\text{for } i\in\mathbb{N} \label{eq:lp:MonotonicitySampling}\\
	C_i+C_j &\ \geq \ C_{i+j}  &\text{for } i,j\in\mathbb{N} 
	\label{eq:lp:TriangleInequalitiesSampling}\\
	\sum_{k=1}^\infty (1-\sigma p)^{k-1}\cdot \left( \alpha \cdot C_k + 2p \cdot \sum_{i=1}^{k-1} \min\left\{\big. C_i,\, C_{k-i}\right\} \right) &\ \geq \ \sigma^{-2}.   \label{eq:lp:MasterRoutesSampling}
\end{align}

Note that in \eqref{eq:lp:MasterRoutesSampling} we only require that \eqref{eq:costofsampledmasterroutesolutionsampling} is at least 1 instead of exactly 1. This does not change the infimum because we can always scale all the $C_i$'s. We have proved:

\begin{lemma}\label{lemma:sampling-OP}
	Let $0<p<1$.
	The reciprocal of the value of \eqref{eq:exact_lp_for_samplingalgo}
	is an upper bound on the approximation guarantee for the sampling algorithm with $f(p) = \sigma p$ 
	for all $p$-normalized instances. 
	\hfill\qed
\end{lemma}

\subsection{Obtaining a single linear program} \label{section:sampling_singleLP}

Note that we have an infinite set of optimization problems (one for each choice of $p$), and, 
in view of \cref{lemma:wlog_normalized}, we have to consider the limit for $p\to 0$.

In the following, we require that $p$ is of the form $p = \frac{\beta}{b}$ for some odd integer $b \ge b_0$. Note that $p \to 0$ as $b \to \infty$. In order to obtain a single optimization problem for all such values of $p$,
we put subsequent $C_i$'s into buckets of size $b$.
More precisely, we define buckets 
\begin{equation} \label{eq:definebuckets}
B_i \ \coloneqq \ \sum\limits_{j=\max\{1,ib-\frac{b-1}2\}}^{ib+\frac{b-1}2} C_{j}
\end{equation}
for $i \ge 0$.  In the following, we show that we can use the constraints in \eqref{eq:exact_lp_for_samplingalgo} to generate (slightly relaxed) constraints that only depend on these buckets. 
First, we note that the buckets are chosen such that they still satisfy the triangle inequality.

\vspace*{-2mm}

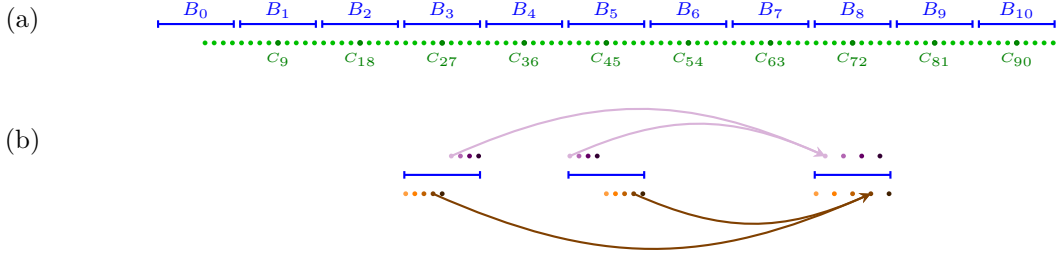
\begin{figure}[ht]
	\begin{center}
		\begin{tikzpicture}[xscale=0.6, thick]
			\tikzstyle{singlevertex}=[darkgreen,circle,draw,minimum size=1,inner sep=0pt]
			\tikzstyle{mainsinglevertex}=[verydarkgreen,circle,draw,minimum size=1.3,inner sep=0pt]
			\tikzstyle{bucket}=[blue]
			\tikzstyle{mainbucket}=[darkblue, line width=1.5]
			\tikzstyle{interval}=[red!85!black, line width=1.2]
			\begin{scope}[shift={(0,0)}]
				\node at (-3, 0.3) {(a)};
				\foreach [evaluate=\i as \pos using (\i / 5)] \i in {5, 6, ..., 98} 
				{
					\node[singlevertex] at (\pos,0) {};
				}
				\foreach 
				[evaluate=\i as \pos using (\i / 5),
				evaluate=\i as \index using (int(\i -4))] 
				\i in {13, 22, ..., 94} 
				{
					\node[mainsinglevertex] at (\pos,0) {};
					\node[verydarkgreen,below] at (\pos, 0) {\tiny$C_{\index}$};
				}
				\foreach 
				[evaluate=\i as \end using (9/5 * \i + 8/5 + 0.03), 
				evaluate=\i as \begin using (9/5 * \i - 0.03)] 
				\i in {0, 1, ..., 10} 
				{
					\draw[bucket] (\begin,0.25) to node[above=-0.5mm]{\scriptsize $B_{\i}$} (\end,0.25);
					\draw[bucket] (\begin,0.2) -- (\begin,0.3);
					\draw[bucket] (\end,0.2) -- (\end,0.3);
				}
			\end{scope}
			\begin{scope}[shift={(0,-2)}]
				\node at (-3, 0.7) {(b)};
				\foreach 
				[evaluate=\i as \end using (9/5 * \i + 8/5 + 0.03), 
				evaluate=\i as \begin using (9/5 * \i - 0.03)] 
				\i in {3,5,8} 
				{
					\draw[bucket] (\begin,0.25) to  (\end,0.25);
					\draw[bucket] (\begin,0.2) -- (\begin,0.3);
					\draw[bucket] (\end,0.2) -- (\end,0.3);
				}
			    \node[singlevertex, orange!75!white] at (27/5, 0) {};
			    \node[singlevertex, orange] at (28/5, 0) {};
			    \node[singlevertex, orange!75!black] at (29/5, 0) {};
			    \node[singlevertex, orange!50!black] at (30/5, 0) {};
			    \node[singlevertex, orange!25!black] at (31/5, 0) {};
			    \node[singlevertex, orange!75!white] at (49/5, 0) {};
			    \node[singlevertex, orange] at (50/5, 0) {};
			    \node[singlevertex, orange!75!black] at (51/5, 0) {};
			    \node[singlevertex, orange!50!black] at (52/5, 0) {};
			    \node[singlevertex, orange!25!black] at (53/5, 0) {};
			    \node[singlevertex, orange!75!white] at (72/5, 0) {};
			    \node[singlevertex, orange] at (74/5, 0) {};
			    \node[singlevertex, orange!75!black] at (76/5, 0) {};
			    \node[singlevertex, orange!50!black] at (78/5, 0) {};
			    \node[singlevertex, orange!25!black] at (80/5, 0) {};
			    
			    \node[singlevertex, violet!30!white] at (32/5, 0.5) {};
			    \node[singlevertex, violet!60!white] at (33/5, 0.5) {};
			    \node[singlevertex, violet!80!black] at (34/5, 0.5) {};
			    \node[singlevertex, violet!40!black] at (35/5, 0.5) {};
			    \node[singlevertex, violet!30!white] at (45/5, 0.5) {};
			    \node[singlevertex, violet!60!white] at (46/5, 0.5) {};
			    \node[singlevertex, violet!80!black] at (47/5, 0.5) {};
			    \node[singlevertex, violet!40!black] at (48/5, 0.5) {};
			    \node[singlevertex, violet!30!white] at (73/5, 0.5) {};
			    \node[singlevertex, violet!60!white] at (75/5, 0.5) {};
			    \node[singlevertex, violet!80!black] at (77/5, 0.5) {};
			    \node[singlevertex, violet!40!black] at (79/5, 0.5) {};
			    \draw[-stealth, violet!30!white] (32/5, 0.5) to[bend left=15] (73/5, 0.5);
			    \draw[-stealth, violet!30!white] (45/5, 0.5) to[bend left=15] (73/5, 0.5);
			    \draw[-stealth, orange!50!black] (30/5, 0) to[bend right=15] (78/5, 0);
			    \draw[-stealth, orange!50!black] (52/5, 0) to[bend right=15] (78/5, 0);
			\end{scope}
			
		\end{tikzpicture}
	\end{center}
	\caption{(a): The green dots stand for $C_1,C_2,\ldots$, and the centers of the buckets ($C_{ib}$ for $i\ge 1$) are highlighted.
	Here the bucket size is $b=9$, and the blue intervals show the buckets $B_0,B_1,B_2,\ldots$
	(b): Combining the triangle inequalities for the $C_i$'s leads to triangle inequalities for the $B_i$'s; here shown for $B_i = B_3$ and $B_j = B_5$: We add up all triangle inequalities for $C_k$ from $B_3$ and $C_{\ell}$ from $B_5$ where $C_k$ and $C_{\ell}$ have the same color; illustrated with $C_{26} + C_{48} \le C_{74}$ and $C_{28} + C_{41} \le C_{69}$.
	\label{fig:bucket_triangle_ineq}}
\end{figure}

\begin{proposition}\label{prop:triangle_buckets}
	For all $i,j \ge 1$,
	\begin{equation*}
		B_{i+j} \ \le \ B_i + B_j .
	\end{equation*}
\end{proposition}	
\begin{proof}
	Indeed, using \eqref{eq:lp:TriangleInequalitiesSampling}, as illustrated in \cref{fig:bucket_triangle_ineq},
	\begin{align*}
		B_{i+j} &\ = \ \sum_{k = - \frac{b-1}2}^{\frac{b-1}2} C_{(i+j)b+k}
		\ = \ \sum\limits_{k=0}^{\frac{b-1}{2}} C_{(i+j)b - \frac{b-1}2 + 2k}
		+ \sum\limits_{k=1}^{\frac{b-1}{2}} C_{(i+j)b - \frac{b-1}2 + 2k-1} \\
		&\ \le \ \sum\limits_{k=0}^{\frac{b-1}{2}} \left(C_{ib - \frac{b-1}2 +k} + C_{jb+k} \right)
		+ \sum\limits_{k=1}^{\frac{b-1}{2}} \left(C_{ib+k} + C_{jb - \frac{b-1}2+k-1}\right) \\
		&\ = \ \sum_{k = -\frac{b-1}2}^{\frac{b-1}2} C_{ib+k} + \sum_{k = -\frac{b-1}2}^{\frac{b-1}2} C_{jb+k}
		\ = \ B_i + B_j .\qedhere
	\end{align*}
\end{proof}

\vspace*{-2mm}

Next we aim for an upper bound on the left-hand side of \eqref{eq:lp:MasterRoutesSampling} that only depends on the buckets. 
First we show:

\vspace*{-2mm}

\begin{lemma} \label{lemma:perfectmatching}
\[
\sum_{k=1}^\infty (1-\sigma p)^{k-1}\cdot  \sum_{i=1}^{k-1} \min\left\{\big. C_i,\, C_{k-i}\right\} \ \le \ b \cdot  \sum_{i=0}^\infty\sum_{j=0}^\infty e^{-(i+j-1)\cdot \sigma bp}\cdot \min\{B_i,B_j\}.
\]	
\end{lemma}

\begin{proof}
	For $i \in \mathbb{Z}_{\ge 0}$, let $I_i = \{\max\{1,ib - \frac{b-1}2\},\ldots, ib + \frac{b-1}2\}$ be the set of indices in the $i$-th bucket. Then
\begin{align*}
	&\ \sum_{k=1}^\infty (1-\sigma p)^{k-1}\cdot  \sum_{i=1}^{k-1} \min\left\{\big. C_i,\, C_{k-i}\right\} \\
	&\ = \ \sum_{k=1}^\infty \sum_{\ell=1}^\infty (1-\sigma p)^{k+\ell-1}\cdot   \min\left\{\big. C_k,\, C_\ell\right\} \\
	&\ \le \ \sum_{i=0}^\infty\sum_{j=0}^\infty (1-\sigma p)^{\max\{0,i+j-1\}\cdot b}\cdot\sum_{k \in I_i}\sum_{\ell\in I_j}\min\{C_k,C_\ell\} \\
	&\ \le \ \sum_{i=0}^\infty\sum_{j=0}^\infty e^{-(i+j-1)\cdot \sigma bp}\cdot\sum_{k \in I_i}\sum_{\ell\in I_j}\min\{C_k,C_\ell\}.
\end{align*}	
In the last inequality we used $1-x \le e^{-x}$ for all $x \in \mathbb{R}$.
Now, for $i,j \in \mathbb{Z}_{\ge 0}$, consider the complete bipartite graph $H$ where one bipartition consists of $|I_j|$ copies of every element of $I_i$, and the other bipartition consists of $|I_i|$ copies of every element of $I_j$. Then 
\[\sum_{(k,\ell)\in E(H)}\min\{C_k,C_\ell\} \ = \ |I_i|\cdot|I_j|\cdot \sum_{k\in I_i}\sum_{\ell\in I_j} \min\{C_k,C_\ell\}.\]
We can partition $E(H)$ into $t\coloneqq |I_i|\cdot |I_j|$ perfect matchings $M_1,\dots,M_t$. Then
\begin{align*}
	 \sum_{(k,\ell)\in E(H)}\min\{C_k,C_\ell\}&\ = \ \sum_{s=1}^t\sum_{(k,\ell)\in M_s}\min\{C_k,C_\ell\}\\
	 &\ \leq \ \sum_{s=1}^t\min\left\{\sum_{(k,\ell)\in M_s}C_k,\sum_{(k,\ell)\in M_s}C_\ell\right\} \\
	 &\ = \ \sum_{s=1}^t\min\left\{|I_j|\cdot \sum_{k\in I_i} C_k, |I_i|\cdot\sum_{\ell\in I_j} C_\ell\right\}\\
	 &\ = \ |I_i|\cdot |I_j|\cdot\min\left\{|I_j|\cdot B_i, |I_i|\cdot B_j\right\} .
\end{align*}
Note that the second equality follows from the fact that $V(H)$ contains $|I_j|$ copies of each element in $I_i$ and vice versa. Moreover, summing over the endpoints of the edges in a perfect matching in $M$ is the same as summing over $V(H)$.
Division by $|I_i|\cdot |I_j|$ yields
\[\sum_{k\in I_i}\sum_{\ell\in I_j} \min\{C_k,C_\ell\} \ \leq \ \min\left\{|I_j|\cdot B_i, |I_i|\cdot B_j\right\} \ \leq \ b \cdot\min\{B_i,B_j\} . \qedhere \]
\end{proof}

Using \cref{lemma:perfectmatching} and $\beta=bp$, 
the left-hand side of \eqref{eq:lp:MasterRoutesSampling} can be upper bounded by 
\begin{align}\label{eq:infinite_many_buckets}
	& \alpha \cdot \sum_{k=1}^\infty (1-\sigma p)^{k-1} \cdot C_k  \ + \ 2 bp \cdot  \sum_{i=0}^\infty\sum_{j=0}^\infty e^{-(i+j-1)\cdot \sigma bp}\cdot \min\{B_i,B_j\} \notag \\
	&\le \ \alpha \cdot \sum_{k=0}^\infty (1-\sigma p)^{\max\{0,kb-\frac{b-1}2-1\}}\cdot B_k  \ + \ 2 bp \cdot  \sum_{i=0}^\infty\sum_{j=0}^\infty e^{-(i+j-1)\cdot \sigma bp}\cdot \min\{B_i,B_j\} \notag \\
	&\leq \ \alpha \cdot \sum_{k=0}^\infty e^{-(k-1) \cdot \sigma \beta}\cdot B_k  \ + \ 2 \beta \cdot  \sum_{i=0}^\infty\sum_{j=0}^\infty e^{-(i+j-1)\cdot \sigma \beta}\cdot \min\{B_i,B_j\} 
	.
\end{align}
The last inequality follows from $(1-\sigma p)^{kb-\frac{b-1}{2}-1}\leq (1-\sigma p)^{kb-b}$ and $1+x\leq e^x$ for all $x\in\mathbb{R}$.
Note that we still sum over infinitely many variables. 
Hence, in order to get a finite linear program, we aim for an upper bound on \eqref{eq:infinite_many_buckets} 
that only depends on the buckets $B_i$ with $i \le N$ for some integer $N$ that we will choose later. 
For this, we use the triangle inequality (\cref{prop:triangle_buckets}) 
to bound the terms depending on buckets $B_i$ with $i > N$ by some term depending on $B_1,\dots,B_N$ only. 
For large $N$ this will result in a negligible error as the coefficients in \eqref{eq:infinite_many_buckets} decrease exponentially.
In \cref{section:proofofboundondelta} we prove the following bound on the error term:

\begin{lemma}\label{lemma:bounddelta}
Let $\delta_1 \coloneqq \frac{4\beta}{e^{N\sigma\beta}(e^{\sigma\beta}-1)}$  and $\delta_2 \coloneqq \left(\alpha + \frac{2\beta}{e^{N\sigma\beta}(e^{\sigma\beta}-1)} \right)\cdot \frac{e^{-N\sigma\beta}}{(1-e^{-\sigma\beta})^2}\cdot (1+N-e^{-\sigma\beta}N)$.
Then
\begin{align*}
 &\alpha \cdot \sum_{k=N+1}^\infty e^{-(k-1) \cdot \sigma \beta}\cdot B_k  \ + \ 2 \beta \cdot  \sum_{i,j \in \mathbb{Z}_{\ge 0}: \max\{i,j\} > N} e^{-(i+j-1)\cdot \sigma \beta}\cdot \min\{B_i,B_j\} \\
 &\ \le \ \delta_1 \cdot  \sum_{k=0}^N  e^{-(k-1)\cdot \sigma \beta}\cdot B_i  + \delta_2 \cdot B_1  
 .  
\end{align*}
\end{lemma}

Therefore, we get a lower bound on  \eqref{eq:exact_lp_for_samplingalgo} by 
minimizing $\sum_{i=1}^{\infty} (1-p)^{i-1}\cdot C_i$ subject to \eqref {eq:definebuckets} and $B_i\ge 0$ for $i\ge 0$, $B_{i+j}\le B_i+B_j$ for $i,j\ge 1$ with $i+j \le N$, and 
\begin{align}\label{eq:sampling_final_mr_constraint}
\!\!\!\!\!\!\!\!\!\! (\alpha + \delta_1) \cdot \sum_{k=0}^N e^{-(k-1) \cdot \sigma \beta}\cdot B_k  + 4 \beta \cdot  \sum_{j=0}^N\sum_{i=0}^j e^{-(i+j-1)\cdot \sigma \beta}\cdot \min\{B_i,B_j\}
	+ \delta_2 \cdot B_1 \geq \sigma^{-2} .
\end{align}

Note that the objective still contains infinitely many variables and depends on $p$.
The first problem can easily be resolved by bounding 
\begin{align}\label{eq:lb_obj}
	\sum_{i=1}^{\infty} (1-p)^{i-1}\cdot C_i 
	\ \ge \ \sum_{i=0}^{\infty}  (1-p)^{bi+\frac{b-1}2-1}\cdot B_i
	\ \ge \ \sum_{i=0}^{N} (1-p)^{(i+\frac12)b}\cdot B_i.
\end{align}
It remains to get rid of the dependence on $b$ and $p$ (recall that $p=\frac{\beta}{b}$). 
To this end, we exploit that $\lim_{b\rightarrow\infty}(1-\frac{\beta}{b})^{(i+\frac12)b}=e^{-(i+\frac12)\beta}$ for all $i=0,\dots,N$, 
and that by \cref{lemma:wlog_normalized}, we can choose $b_0$ arbitrarily large. 
This will allow us to conclude that we can replace the objective by $\sum_{i=0}^{N} e^{-(i+\frac12)\beta}\cdot B_i$ and still obtain an upper bound (see the proof of \cref{lem:primal_lp_sampling} for the technical details).
Putting everything together, we arrive at the following LP. 

\vspace*{-4mm}
{\small
\begin{align}
	\min \sum_{i=0}^{N} e^{-(i+\frac12)\beta}\cdot B_i  &&  \tag{Sampling-LP} \label{eq:ub_lp_for_samplingalgo} \\[-1mm]
	\hspace*{-2mm} \text{subject to } \hspace*{2.2cm} & \hspace*{-2cm}
	(\alpha + \delta_1) \cdot \sum_{k=0}^N e^{-(k-1) \cdot \sigma \beta}\cdot B_k + \delta_2 \cdot B_1  
	&\hspace*{-10mm} + \  4 \beta \cdot  \sum_{j=0}^N\sum_{i=0}^j e^{-(i+j-1)\cdot \sigma \beta}\cdot M_{i,j} 
	\ \geq \ \sigma^{-2} \hspace*{-5mm} & \label{eq:lp:MasterRoutesSamplingFinal} \\
	B_i+B_j &\ \geq \ B_{i+j}  &\text{for } 1 \le i \le j \le N \text{ with } i+j\leq N
	\label{eq:lp:TriangleInequalitiesSamplingFinal}\\
	B_i &\ \ge \ M_{i,j} &\text{for } 0 \le i \le j \le N \label{eq:lp:MVars1SamplingFinal} \\ 
	B_j &\ \ge \ M_{i,j} &\text{for } 0 \le i \le j \le N \label{eq:lp:MVars2SamplingFinal} \\
	B, M &\ \ge \ 0.
\end{align}
}
Recall that $\delta_1$ and $\delta_2$ were defined in \cref{lemma:bounddelta}.
We conclude:

\begin{lemma}\label{lem:primal_lp_sampling}
	Let $N$ be an integer and $\beta>0$.
	The reciprocal of the optimum value of \eqref{eq:ub_lp_for_samplingalgo}
	is an upper bound on the approximation guarantee of the sampling algorithm for 
	$f(p)=1-(1-p)^\sigma$ (using an $\alpha$-approximation algorithm for TSP), for all \textsc{a priori TSP} instances with depot.
\end{lemma}

\begin{proof}
	We compare the value of \eqref{eq:ub_lp_for_samplingalgo} to the value of \eqref{eq:exact_lp_for_samplingalgo}.
	We showed above that for any feasible solution $C$ to \eqref{eq:exact_lp_for_samplingalgo} 
	we obtain a feasible solution $(B,M)$ to \eqref{eq:ub_lp_for_samplingalgo}
	via \eqref{eq:definebuckets} and $M_{i,j}=\min\{B_i,B_j\}$. 
	
	Fix $\delta>0$. 
	Then there exists $b_0\in\mathbb{N}$ such that for all odd integers $b\ge b_0$
	\begin{equation*}
		\sum_{i=0}^{N} e^{-(i+\frac12)\beta} \cdot B_i
		\ \le \ (1+\delta)\cdot\sum_{i=0}^{N}\left(1-\textstyle{\frac{\beta}{b}}\right)^{(i+\frac12)b} \cdot B_i
		\ \stackrel{\eqref{eq:lb_obj}}{\le} \ (1+\delta)\cdot\sum_{i=0}^{\infty}\left(1-\textstyle{\frac{\beta}{b}}\right)^{i-1} \cdot C_i .
	\end{equation*} 
	Thus the value of  \eqref{eq:ub_lp_for_samplingalgo} is at most $(1+\delta)$ times the value of \eqref{eq:exact_lp_for_samplingalgo}
	with $p=\frac{\beta}{b}$ for all odd integers $b\ge b_0$.
	Hence, by \cref{lemma:sampling-OP}, $(1+\delta)$ times the reciprocal of the optimum value of 
	\eqref{eq:ub_lp_for_samplingalgo} is an upper bound on \eqref{eq:apxratio_of_sampling} 
	for all $\frac{\beta}{b}$-normalized instances for all odd integers $b\ge b_0$. 
	By \cref{lemma:wlog_normalized}, the same bound then holds for all instances with depot.
	Since this bound holds for all $\delta>0$, it also holds for $\delta=0$.
\end{proof}

\subsection{The dual LP}

In order to obtain a lower bound on the optimum value of \eqref{eq:ub_lp_for_samplingalgo}, 
we provide a \emph{feasible solution} to the \emph{dual} linear program.
For the dual LP, we introduce variables $x_{i,j}$ for the inequalities of type \eqref{eq:lp:TriangleInequalitiesSamplingFinal}, variables $v_{i,j}$ and $w_{i,j}$ for the inequalities of type \eqref{eq:lp:MVars1SamplingFinal} and \eqref{eq:lp:MVars2SamplingFinal}, respectively, and a variable $y$ for inequality \eqref{eq:lp:MasterRoutesSamplingFinal}.
Using these variables, the dual LP looks as follows:
\begin{small}
\begin{align}
\max \ \sigma^{-2}\cdot y && \tag{Dual-Sampling-LP}\label{eq:DualLPSamplingFinal} &&\\
\text{subject to} \hspace{2cm} 4 \beta\cdot e^{-(i+j-1)\cdot \sigma \beta}\cdot y &\ \le \  v_{i,j}+w_{i,j} &\!\!\!\!\!\!\!\!\!\!\!\text{for } 0\leq i\le j\leq N  \label{eq:lp:DualSamplingMvars}\\
\!\!\!\!\!\!\!\!\! (\alpha + \delta_1) \cdot e^{-(k-1) \cdot \sigma \beta}\cdot y+\sum_{j=k}^{N} v_{k,j}+ \sum_{j=0}^k w_{j,k} + \mathds{1}_{k=1} \cdot \delta_2 \cdot y  &&\notag\\ 
+ \mathds{1}_{k > 0} \cdot \left( \sum_{i=1}^{\min\{k,N-k\}}x_{i,k}+\sum_{i=k}^{N-k} x_{k,i}-\sum_{\substack{1\leq i\leq j\leq N,\\i+j = k}}x_{i,j} \right) & \ \leq \ e^{-(k+\frac12)\beta} & \text{for } 0\leq k\leq N  \label{eq:lp:DualSamplingCjFinal}\\
 x,y,v,w & \ \ge \ 0. & 
\end{align}
\end{small}

\vspace*{-5mm}
\begin{corollary}
	Let $N$ be an integer and $\beta>0$.
	For any feasible solution $(x,y,v,w)$ to \eqref{eq:DualLPSamplingFinal}, $\sfrac{\sigma^2}{y}$ is an upper bound on the approximation ratio of the sampling algorithm with $f(p)=1-(1-p)^\sigma$ restricted to \textsc{a priori TSP} instances with depot. 
\hfill\qed
\end{corollary}	

We have computed a dual solution using Gurobi 10.0.1 with $\alpha=1.5$, $\beta = \frac1{100}$, $N = 2500$, 
and $\sigma=0.663$, yielding an upper bound of $3.094$ and thus proving \cref{thm:ub_sampling}. 
The dual solution and a Python script that verifies that this is a feasible solution to \eqref{eq:DualLPSamplingFinal} can be found at \url{https://doi.org/10.60507/FK2/JCUIRI}. For $\alpha = 1$, we get an upper bound of $2.694$.

\begin{comment_}\label{comment:conjecture_sampling}
Solving \eqref{eq:ub_lp_for_samplingalgo} with the same values for $\alpha$, $\beta$, $N$, and $\sigma$ 
yields an \textsc{a priori TSP} instance of the same shape as the example provided in \cref{sec:sampling_lb}. 
Hence we conjecture 
that the upper bound given by \eqref{eq:DualLPSamplingFinal} converges to the lower bound given in \cref{thm:sampling_lb} 
for $\beta\to 0$ and $N\beta \rightarrow \infty$.
\end{comment_}

\subsection{Bounding the error term (Proof of \cref{lemma:bounddelta})} \label{section:proofofboundondelta}
We first prove the following auxiliary lemma:

\begin{lemma}\label{lem:auxiliary}
	Let $n \in \mathbb{N}$ and $q \in (0,1)$. Then
	\begin{equation} \label{eq:ind_hypo}
		\sum_{k=n+1}^\infty k\cdot q^{k-1} \ = \ \frac{q^{n}}{(1-q)^2}\cdot(1+n-q n).
	\end{equation}
\end{lemma}
\begin{proof}
	By induction on $n$. For $n=0$, the statement is equivalent to the well-known formula 
	\[\sum_{k=1}^\infty k\cdot (1-q)\cdot q^{k-1} \ = \ \frac{1}{1-q}\] 
	for the expected value of a geometrically distributed random variable. 
	Next, assume that \eqref{eq:ind_hypo}
	holds for some $n\in\mathbb{N}$. Then
	\begin{align*}
		&\sum_{k=n+2}^\infty k\cdot q^{k-1} \ = \ \sum_{k=n+1}^\infty k\cdot q^{k-1} - (n+1)\cdot q^n \stackrel{\eqref{eq:ind_hypo}}{\ = \ } \frac{q^{n}}{(1-q)^2}\cdot(1+n-q n) -(n+1)\cdot q^n\\
		&\ = \ \frac{q^{n}}{(1-q)^2}\cdot(1+n-q n - (n+1)\cdot (1-q)^2) \ = \ \frac{q^{n+1}}{(1-q)^2}\cdot(n+2-q (n+1)),
	\end{align*}
	which is \eqref{eq:ind_hypo} for $n+1$.
\end{proof}
Now we are ready to prove \cref{lemma:bounddelta}:
\begin{proof}[Proof of \cref{lemma:bounddelta}]
	We compute
	\begin{align*}
		& 2 \beta \cdot  \sum_{i,j \in \mathbb{Z}_{\ge 0}: \max\{i,j\} > N} e^{-(i+j-1)\cdot \sigma \beta}\cdot \min\{B_i,B_j\}  \\
		&\le \ 
		4 \beta \cdot  \sum_{i=0}^N \sum_{j=N+1}^{\infty} e^{-(i+j-1)\cdot \sigma \beta}\cdot B_i  
		\ + \ 2\beta \cdot  \sum_{i=N+1}^\infty \sum_{j=N+1}^\infty e^{-(i+j-1)\cdot \sigma \beta } \cdot B_i \\
		&= \  
		4 \beta \cdot  \sum_{i=0}^N  e^{-(i-1)\cdot \sigma \beta}\cdot B_i \cdot \sum_{j=N+1}^{\infty} e^{-j\sigma\beta}   
		\ + \ 2\beta \cdot  \sum_{i=N+1}^\infty e^{-(i-1)\cdot \sigma \beta } \cdot B_i \cdot \sum_{j=N+1}^\infty e^{-j \sigma \beta } \\
		&= \ 
		\delta_1 \cdot  \sum_{i=0}^N  e^{-(i-1)\cdot \sigma \beta}\cdot B_i  
		\ + \ \frac{2\beta}{e^{N\sigma\beta}(e^{\sigma\beta}-1)} \cdot  \sum_{i=N+1}^\infty e^{-(i-1)\cdot \sigma \beta } \cdot B_i \\
	\end{align*}
	Bounding $B_i \le i\cdot B_1$ for $i > N$ by using to the triangle inequality (\cref{prop:triangle_buckets}), we obtain
	\begin{align*}
		\!\!&\alpha \cdot \sum_{k=N+1}^\infty e^{-(k-1) \cdot \sigma \beta}\cdot B_k  \ + \ 
		2 \beta \cdot  \sum_{i,j \in \mathbb{Z}_{\ge 0}: \max\{i,j\} > N} e^{-(i+j-1)\cdot \sigma \beta}\cdot \min\{B_i,B_j\}  \\
		\!\!&\le \ \delta_1 \cdot  \sum_{i=0}^N  e^{-(i-1)\cdot \sigma \beta}\cdot B_i  +
		\left(\alpha + \ \frac{2\beta}{e^{N\sigma\beta}(e^{\sigma\beta}-1)} \right)\cdot \sum_{k=N+1}^\infty e^{-(k-1) \cdot \sigma \beta}\cdot B_k  \\
		\!\!&\le \ \delta_1 \cdot  \sum_{i=0}^N  e^{-(i-1)\cdot \sigma \beta}\cdot B_i  +
		\left(\alpha + \ \frac{2\beta}{e^{N\sigma\beta}(e^{\sigma\beta}-1)} \right)\cdot \sum_{k=N+1}^\infty e^{-(k-1) \cdot \sigma \beta}\cdot k\cdot B_1  \\
		\!\!&= \ \delta_1 \cdot  \sum_{i=0}^N  e^{-(i-1)\cdot \sigma \beta}\cdot B_i  + \delta_2 \cdot B_1  
		,  
	\end{align*}
	where we used \cref{lem:auxiliary} in the final equality with $n =N$ and $q = e^{-\sigma\beta}$.
\end{proof}

\section{Lower bound on the approximation ratio of the sampling algorithm}
\label{sec:sampling_lb}

In this section we prove \cref{thm:sampling_lb}.
We will provide a family of instances for which the sampling algorithm that we described in \cref{sec:our_results} has no better approximation ratio than $2.655$,
no matter how we choose $f$, and even when assuming that we can compute optimal TSP tours on the sampled customers. 
Using the currently best approximation guarantee for metric TSP leads to a ratio of more than $3.049$ (again, even if we try all $f$).

In the proof we exploit that each instance in the family that we describe has uniform activation probability, 
i.e., $p(v) = p$ for each $v \in V$, where $p$ is a small positive number. 
Hence, it suffices to consider functions $f$ with $f(1)=1$ and $f(p)=\sigma p$ for some $\sigma \in \left[0, \frac{1}{p}\right]$.

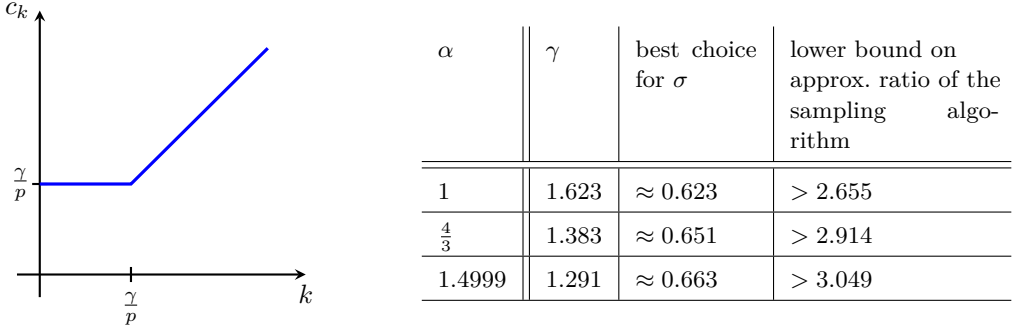
\begin{figure}[h]
	\begin{minipage}{0.3\textwidth}
		\begin{tikzpicture}[thick]
			\draw[-stealth] (-0.3, 0) to node[below, pos=1]{$k$} (3.5,0);
			\draw[-stealth] (0, -0.3) to node[left, pos=1]{$c_k$} (0,3.5);
			\draw (1.2, -0.1) to node[below=1mm]{$\frac{\gamma}p$} (1.2, 0.1);
			\draw (-0.1, 1.2) to node[left]{$\frac{\gamma}{p}$} (0.1, 1.2);
			\draw[very thick, blue] (0,1.2) -- (1.2,1.2) -- (3, 3);
		\end{tikzpicture}
	\end{minipage}
	\hfill
	\begin{minipage}{0.6\textwidth}
		\renewcommand{\arraystretch}{1.25}
		\begin{tabular}{l||l|p{1.6cm}|p{2.7cm}}
			$\alpha$ & $\gamma$ & best choice for $\sigma$& \mbox{lower bound on} \mbox{approx.\ ratio of the} sampling algorithm\\\hline\hline
			1 & $1.623$ & $\approx 0.623$ & $>2.655$ \\\hline
			$\frac43$ & $1.383$ & $\approx 0.651$ & $>2.914$ \\\hline
			1.4999 & 1.291 & $\approx 0.663$ & $> 3.049$ \\\hline
		\end{tabular}
	\end{minipage}
	\caption{ \label{fig:lb_ex_sampling} Left: The distance of $v_i$ and $v_j$ for $0\leq i<j\leq n-1$ is $\frac{c_k}{n}$, depending on $k = \min\{j-i, i+n-j\}$. 
	The figure shows the dependence of $c_k$ on $k$.
		Right: The table shows, for a given approximation ratio $\alpha$ that our black box TSP approximation algorithm achieves, 
		how we should choose $\gamma$ such that the sampling algorithm performs worst possible even if we choose $\sigma$ best possible. 
	}
\end{figure}

Let $\gamma\in [1,2]$ be a parameter chosen later, depending only on the approximation ratio $\alpha$ for TSP. 
Let $0 < \varepsilon \ll 1$.
We will choose $p>0$ (very small) and $n\in\mathbb{N}$ (very large), depending on $\gamma$ and $\epsilon$ only (cf.\ \cref{lem:choice_of_p_and_n}).
We then consider a $p$-normalized instance of the \textsc{a priori TSP} with $V=\{v_0,\cdots,v_{n-1}\}$ with $v_0$ being the depot.
Then for $0\le i < j \le n-1$ with $k=\min\{j-i,\, i+n-j\}$ define distances $c(v_i,v_j)=\frac{c_k}{n}$, where (cf.\ \cref{fig:lb_ex_sampling})
\begin{equation*}
	c_{k} \ \coloneqq \ \begin{cases}
		\frac{\gamma}{p} & \textnormal{if } k \le \frac{\gamma}{p}\\
		k & \textnormal{otherwise.}\\
	\end{cases}
\end{equation*}
Note that they form a metric. We will show the claimed lower bounds for this set of instances.

We choose $p$ and $n$ such that the following properties hold that we will use later:

\begin{lemma}\label{lem:choice_of_p_and_n} 
For every $\gamma \in [1,2]$ and every $\epsilon\in (0,\frac{1}{4})$, there are $p>0$ and $n\in\mathbb{N}$ such that $\frac{\gamma}{p}$ is integral and 
	\begin{enumerate}
		\item\label{item:small_p} $p \le (1-4\epsilon)\cdot\epsilon^2$,
		\item\label{item:small_p_two} $(1-\varepsilon p)\cdot(1-xp)^{\frac{y}{p}} \ge e^{-x y}-\varepsilon$ for all $x \in \big[0,\frac1{p}\big]$ and $y \in \big[1,\frac{2}{\varepsilon}\big]$,
		\item\label{item:large_n} $n \ge \frac{4}{\varepsilon p} + \frac{3}{\varepsilon}$,
		\item\label{item:large_N} $(1+x)\cdot e^{-p\varepsilon^2x} \le \varepsilon-\frac{1}{n}$ for all $x \ge \varepsilon^2pn -3$.
	\end{enumerate}
\end{lemma}
\begin{proof}
Fix $\gamma\in [1,2]$ and $\varepsilon \in (0,\frac{1}{4})$. As $\lim_{t\to 0} (1-t)^{-\frac{2\ln(\varepsilon)}{\varepsilon}} = 1$, pick $\delta\in (0,1)$ with
\begin{equation}
1-\frac{\varepsilon}{2} \ \leq \ (1-\delta)^{-\frac{2\ln(\varepsilon)}{\varepsilon}} \label{eq:choice_delta}.
\end{equation}
Moreover, as $\lim_{t\to 0}(1-t)^\frac{1}{t}=e^{-1}$, we can pick $p\in (0,\frac{1}{2})$ such that \ref{item:small_p} holds, $\frac{\gamma}{p}$ is integral and such that
\[\forall t\in (0,-\ln(\varepsilon)\cdot p]: \ (1-t)^{\frac{1}{t}} \ \geq \ (1-\delta)\cdot e^{-1}.\]
In particular,
\begin{equation}
\forall x\in (0,-\ln(\varepsilon)]: \ (1-xp)^{\frac{1}{xp}} \ \geq \ (1-\delta)\cdot e^{-1}\label{eq:choice_p}.
\end{equation}
We show that \ref{item:small_p_two} holds. First of all, if $x>-\ln(\varepsilon)$, then the right-hand-side of \ref{item:small_p_two} is negative, whereas the left-hand-side is nonnegative, so we are done in this case. Next, assume that $x\leq -\ln(\varepsilon)$ holds. If $x=0$, then \ref{item:small_p_two} is equivalent to $1-\epsilon p \geq 1-\epsilon$. For $x> 0$, we calculate
\begin{align*}
(1-\varepsilon p)\cdot(1-xp)^{\frac{y}{p}} &\ = \ (1-xp)^{\frac{y}{p}}-\varepsilon p\cdot (1-xp)^{\frac{y}{p}} &&\mid p\in\left(0,\frac{1}{2}\right)\\
&\ \geq \ \left((1-xp)^{\frac{1}{xp}}\right)^{xy}-\frac{\varepsilon}{2}&&\mid \eqref{eq:choice_p}\\
&\ \geq \ (1-\delta)^{xy}\cdot e^{-xy}-\frac{\varepsilon}{2} &&\mid x\leq -\ln(\varepsilon)\\
&\ \geq \ (1-\delta)^{-\frac{2\ln(\varepsilon)}{\epsilon}}\cdot e^{-xy}-\frac{\varepsilon}{2} &&\mid \eqref{eq:choice_delta}\\
&\ \geq \ \left(1-\frac{\varepsilon}{2}\right)\cdot e^{-xy}-\frac{\varepsilon}{2}\\
&\ \geq \ e^{-xy}-\varepsilon.
\end{align*}
Finally, as $\lim_{x\rightarrow\infty} (1+x)\cdot e^{-p\varepsilon^2 x}=0$, we may choose $n$ subject to \ref{item:large_n} and \ref{item:large_N}.
\end{proof}

To prove \cref{thm:sampling_lb} we use the following auxiliary lemma multiple times:

\begin{lemma}\label{claim:auxiliary}
	For $\beta \in \left(0,\frac{1}{p}\right]$ and $k \in \mathbb{N}_{\ge \frac{\gamma}{p}}$ we have
	\begin{equation}
		\sum_{i=1}^{k} (\beta p)^2 \cdot (1-\beta p)^{i-1} \cdot c_i \ = \ \beta\gamma + (1-\beta p)^{\frac{\gamma}{p}} - (1+ \beta p k) (1-\beta p)^{k},
	\end{equation}
	where $0^0\coloneqq 1$.
\end{lemma}	

\begin{proof}
	The case $\beta p=1$ is straightforward. Next, assume $\beta \in (0,\frac{1}{p})$. For $k=\frac{\gamma}{p}$, we have 
	\begin{align*}
	&\sum_{i=1}^{\frac{\gamma}{p}} (\beta p)^2 \cdot (1-\beta p)^{i-1} \cdot c_i \ = \ \sum_{i=1}^{\frac{\gamma}{p}} (\beta p)^2 \cdot (1-\beta p)^{i-1} \cdot \frac{\gamma}{p} \ = \ (\beta \gamma)\cdot \sum_{i=1}^{\frac{\gamma}{p}}(\beta p) \cdot (1-\beta p)^{i-1}\\
	&\ = \ (\beta \gamma)\cdot (1-(1-\beta p)^{\frac{\gamma}{p}}) \ = \ \beta\gamma + (1-\beta p)^{\frac{\gamma}{p}} - \left(1+ \beta p \cdot\frac{\gamma}{p}\right) (1-\beta p)^{\frac{\gamma}{p}}.
	\end{align*}
	For the case $k\geq \frac{\gamma}{p}+1$, we use \cref{lem:auxiliary} to compute
	\begin{align*}
	&\sum_{i=1}^{k} (\beta p)^2 \cdot (1-\beta p)^{i-1} \cdot c_i\\
	&\ = \ \sum_{i=1}^{\frac{\gamma}{p}} (\beta p)^2 \cdot (1-\beta p)^{i-1} \cdot c_i+(\beta p)^2 \cdot\left(\sum_{i=\frac{\gamma}{p}+1}^\infty  i\cdot (1-\beta p)^{i-1}-\sum_{i=k+1}^\infty  i\cdot (1-\beta p)^{i-1}\right)\\
	&\ = \ (\beta \gamma)\cdot (1-(1-\beta p)^{\frac{\gamma}{p}})+(1-\beta p)^{\frac{\gamma}{p}}\cdot \left(1+\frac{\gamma}{p}\beta p\right)-(1-\beta p)^{k}\cdot (1+k\beta p )\\
	&\ = \ \beta\gamma + (1-\beta p)^{\frac{\gamma}{p}} - (1+ \beta p k) (1-\beta p)^{k}.\qedhere
	\end{align*}
\end{proof}

Now we proceed to the main part of the proof of \cref{thm:sampling_lb}. 
First, we aim for an upper bound on the expected cost of an optimum a priori tour:
\begin{lemma}\label{prop:ub_opt}
	The optimum a priori tour for the considered instance has expected cost at most $(1+\varepsilon) \cdot (\gamma + e^{-\gamma})$.
\end{lemma}
\begin{proof}
	Consider the a priori tour $T^*$ that visits $v_0,\ldots,v_{n-1}$ in this order. This a priori tour has expected cost  
	\begin{align*}
		& \sum_{i=1}^{n-2} \sum_{j=1}^{n-1-i} p^2 \cdot (1-p)^{i-1} \cdot c(v_j,v_{j+i}) + \sum_{i=1}^{n-1} p \cdot (1-p)^{i-1} \cdot \left(\big. c(v_0,v_i) + c(v_{n-i},v_0) \right) \\
		& \le \ \sum_{i=1}^{n-2} p^2 \cdot (1-p)^{i-1} \cdot c_i + 2 \sum_{i=1}^{n-1} p \cdot (1-p)^{i-1} \cdot \frac{c_i}{n} \\
		& \le \ (1+\varepsilon) \cdot \sum_{i=1}^{n-1} p^2 \cdot (1-p)^{i-1} \cdot c_i \\
		&\le \ (1+\varepsilon) \cdot (\gamma + (1-p)^{\frac{\gamma}{p}}) \\[1mm]
		& \le \ (1+\varepsilon) \cdot (\gamma + e^{-\gamma}) ,
	\end{align*}
	where we used $\frac{2}{n}\le\epsilon p$ (which follows from \cref{lem:choice_of_p_and_n}~\ref{item:large_n}) in the second inequality 
	and \cref{claim:auxiliary} for $\beta=1$ and $k=n-1$ in the third inequality.
\end{proof}

Second, we aim for a lower bound on the expected cost of the master route solution:
We define $q:V\rightarrow[0,1]$ by setting $q(v)\coloneqq\sigma p$ for $v\in V\setminus\{v_0\}$ and $q(v_0)\coloneqq 1$.
\begin{lemma}\label{prop:lb_sampling}
	The expected cost of the master route solution when sampling each customer $v\in V$ with probability $q(v)$  is at least
	\[ \min\left\{ \frac{e^{-2}}{\epsilon}-1,\ 
	(1-\epsilon)^3\cdot(1-4\epsilon)\cdot \left( \alpha(\sigma\gamma + e^{-\sigma\gamma}) + 2\gamma + \frac{1}{\sigma}e^{-2\sigma\gamma} \right)
	\right\} .\]
\end{lemma}

\begin{proof}

	We distinguish several cases, depending on how large $\sigma$ is.
	
	\noindent\emph{Case 1:} $\sigma\le\varepsilon$.
	
	In this case, we consider the connection cost only. For each $v_i$ with $\big\lceil\frac{1}{\varepsilon p}\big\rceil \le i \le n- \big\lceil\frac{1}{\varepsilon p}\big\rceil$, 
	the probability that no sampled vertex is fewer than $\big\lceil\frac{1}{\varepsilon p}\big\rceil$ hops on $T^*$ away from $v_i$ is at least
	\begin{equation*}
		(1-\sigma p)^{2\left\lceil\frac{1}{\varepsilon p}\right\rceil -1} \ \ge \ (1- \varepsilon p)^{1+\frac{2}{\varepsilon p}} \ \ge \ e^{-2} - \varepsilon,
	\end{equation*}
	where we used \cref{lem:choice_of_p_and_n}~\ref{item:small_p_two} with $x = \varepsilon$ and $y = \frac2{\varepsilon}$.
	In this event, if $v_i$ is active, we have to pay connection cost at least $2\cdot\frac{1}{n}\big\lceil\frac{1}{\epsilon p}\big\rceil$.
	Hence the total connection cost is at least
	\begin{equation*}
		(e^{-2} - \varepsilon) \cdot \left(n+1 - 2\left\lceil\frac{1}{\varepsilon p}\right\rceil\right) \cdot \frac{2p}{n}\left\lceil\frac{1}{\varepsilon p}\right\rceil 
		\ \ge \ (e^{-2} - \varepsilon) \cdot \left(n-1 - \frac{2}{\varepsilon p}\right) \cdot \frac{2}{\varepsilon n} 
		\ \ge \  \frac{e^{-2}}{\epsilon} - 1,
	\end{equation*}
	where we used that $n \ge \frac{4}{\epsilon p} +2$ by \cref{lem:choice_of_p_and_n}~\ref{item:large_n} in the last inequality.

	\noindent\emph{Case 2:} $\sigma > \varepsilon$.
	Let $S$ denote the set of sampled customers, and, if $|S|\geq 2$, let $e_{\max}[S]$ be a longest edge of the tour $T^*[S]$ that we get from $T^*$ by skipping the
	customers that were not sampled. Note that $S$ and $e_{\max}[S]$ are random variables that depend on the sampling.
	
	We want to compute a lower bound on the expected cost of the master tour.
	It is not always true that $T^*[S]$ is an optimum TSP tour for $S$, but almost. Namely, if $|S|=1$, the statement is true, and otherwise,
	we claim that $c(T^*[S])-c(e_{\max}[S])$ is a lower bound on the cost of any TSP tour for $S$. This follows from
	
	\begin{claim} \label{claim:mst}
	For every $\{v_0\}\subsetneq S\subseteq V$, $T^*[S]\setminus \{e_{\max}[S]\}$ is a min-cost spanning tree in $(S,c)$.
	\end{claim}
\begin{claimproof}
To prove this, we may assume (by the cyclic symmetry of $c$) that $S=\{v_i:i\in I\}$ and $e_{\max}[S] = \{v_{\min I},v_{\max I}\}$; 
then let $i,\ell\in I$ with $i<\ell$, and let $\{j,k\}$ be any edge on the path from $i$ to $\ell$ in the tree $T^*[S]\setminus \{e_{\max}[S]\}$
(i.e., $i\le j< k\le \ell$ and $j,k\in I$). 
We show $c(v_i,v_\ell)\ge c(v_j,v_k)$, which implies optimality of the spanning tree $T^*[S]\setminus \{e_{\max}[S]\}$ 
(see, e.g., Theorem 6.3 in \cite{kortevygen}).
Indeed, $c(v_i,v_\ell)\ge \frac1{n}  \max\bigl\{\frac{\gamma}{p},\,\min\{\ell-i,\, n+i-\ell\} \bigr\} \ge \frac1{n}  \max \bigl\{ \frac{\gamma}{p},\, \min\{k-j,\, n+{\min I}-\max I\} \bigr\}
\ge \min\{c(v_j,v_k),c(e_{\max}[I])\} \ge c(v_j,v_k)$.
This concludes the proof of \cref{claim:mst}.	
\end{claimproof}

	Next we bound the expected cost of $e_{\max}[S]$, or, to be more precise, \[\sum_{\{v_0\}\subsetneq U\subseteq V}\Prob_{S\sim q}[S=U]\cdot c(e_{max}[U]).\]
	The probability that $T^*[S]$ contains an edge of cost at least $\varepsilon+\frac{1}{n}$ is at most
	\[\sum_{j=0}^{n-1} (1-\sigma p)^{\left(\varepsilon+\frac{1}{n}\right)\cdot n-1} \ = \ n\cdot (1-\sigma p)^{\varepsilon n} \ \le \ n\cdot e^{-\sigma p \varepsilon n} \ \le \ n\cdot e^{- p \varepsilon^2 n} \ \le \ \varepsilon-\frac{1}{n}\]
	by \cref{lem:choice_of_p_and_n}~\ref{item:large_N} (when choosing $x = n$).
	As every edge costs less than $1$, the expected length of $e_{\max}[S]$ is less than $2\varepsilon$.
	
	Therefore, using \cref{claim:mst}, the expected cost of the optimum master tour for $S$ is at least $c(T^*[S])-2\varepsilon$, and the expectation
	cannot increase if we sample every customer, including the depot, with probability $\sigma p$. 
	Hence, writing $v_{i}=v_{i-n}$ for $i=n,\ldots,2n-2$ we get as lower bound  
	\begin{align*}
		\Exp_{S\sim q} \left[ \OptTSP (S,c) \right] &\ \ge \ \Exp_{S\sim q} \left[c(T^*[S])\right] - 2\varepsilon \notag \\
		&\ \geq \ \sum_{i=1}^{n-1} \sum_{j=0}^{n-1} (\sigma p)^2 \cdot (1-\sigma p)^{i-1} \cdot c(v_j,v_{j+i})  - 2\varepsilon \notag \\
		&\ = \ \sum_{i=1}^{n-1} (\sigma p)^2 \cdot (1-\sigma p)^{i-1} \cdot c_i  - 2\varepsilon \notag \\
		&\ = \ \left( \sigma \gamma +  (1- \sigma  p)^{\frac{\gamma}{p}} - (1+\sigma p (n-1)) (1-\sigma p)^{n-1} \right)  - 2\varepsilon,
	\end{align*}
	where we used \cref{claim:auxiliary} for $k=n-1$ and $\beta=\sigma$ (recall $ \varepsilon < \sigma \le \frac{1}{p}$) in the last equation.
	Using $\sigma \ge \varepsilon$, we can apply \cref{lem:choice_of_p_and_n}~\ref{item:large_N} with $x = \sigma (n-1)$ and obtain 
	\[(1+\sigma p (n-1))\cdot (1-\sigma p)^{n-1} \ \le \ (1+\sigma (n-1))\cdot e^{-\sigma p (n-1)} \ \le \ (1+\sigma (n-1))\cdot e^{-p \varepsilon^2 \sigma (n-1)} \ \le \ \varepsilon.\] This yields
	\begin{align}
		\Exp_{S\sim q} \left[ \OptTSP (S,c) \right] \ &\ge \ \sigma \gamma +  (1- \sigma  p)^{\frac{\gamma}{p}} - 3\varepsilon\notag\\
		 &\ge \ \sigma\gamma+e^{-\sigma\gamma}-4\epsilon\notag\\
		 &\ge \ (1-4\epsilon)\cdot (\sigma\gamma+e^{-\sigma\gamma})\label{eq:master_tour_cost}
	\end{align}
		using \cref{lem:choice_of_p_and_n}~\ref{item:small_p_two} for $x=\sigma$ and $y=\gamma$ in the second inequality and $x+e^{-x}\geq 1$ for all $x\in\mathbb{R}$ in the last inequality.

	\noindent\emph{Case 2a:} $\sigma\ge\frac{\varepsilon}{p}$.	
	
	Then we get that \eqref{eq:master_tour_cost} is at least 
	\begin{align*}
		(1-4\epsilon)\cdot\frac{\gamma \varepsilon}{p} \ \ge \ \frac{1}{\varepsilon} \ \ge \ \frac{e^{-2}}{\epsilon} - 1,
	\end{align*}
	where we used $\gamma\geq 1$ and \cref{lem:choice_of_p_and_n}~\ref{item:small_p} in the first inequality.

	\noindent\emph{Case 2b:} $\varepsilon < \sigma < \frac{\varepsilon}{p}$.

	We obtain a lower bound on the connection costs as follows:  
	Let $N\coloneqq\big\lfloor \frac{\varepsilon n - 1}{2}\big\rfloor$. 
	Note that $N\geq \frac{\gamma}p$ since $n \geq \frac{2\gamma}{\varepsilon p} + \frac3{\varepsilon}$ by \cref{lem:choice_of_p_and_n}~\ref{item:large_n}.
	We only consider vertices $v_i$ with $N+1\leq i \leq n-1-N$ 
	and connect them to a sampled customer that is at most $N$ hops away on $T^*$, if such a customer exists. 
	Otherwise, we do not connect them at all.
	This yields a lower bound for the expected cost of connecting the active customers to the master tour of
	\begin{align*}
		&\sum_{j=N+1}^{n-1-N} \sum_{i=1}^N 2p \cdot (1-\sigma p)^{2i-1} \cdot \Prob_{S \sim q}\left[S \cap \{v_{j-i}, v_{j+i}\} \neq \emptyset\right] \cdot c(v_j,v_{j+i}) &\\
		&= \ \frac{n-1-2N}{n} \cdot \sum_{i=1}^N 2p \cdot (1-\sigma p)^{2i-1} \cdot \sigma p(2-\sigma p) \cdot c_i &&\mid N=\left\lfloor \frac{\varepsilon n - 1}{2}\right\rfloor\\
		&\ge \ (1-\varepsilon) \cdot   \sum_{i=1}^N 2p \cdot (1-\sigma p)^{2i-1} \cdot \sigma p(2-\sigma p) \cdot c_i &\\
		&\ge \ (1-\varepsilon) \cdot (2-\sigma p) \cdot (1-\sigma p) \cdot \frac{1}{2\sigma} \sum_{i=1}^N (2\sigma p)^2 \cdot (1-2\sigma p)^{i-1}  \cdot c_i &&\mid \sigma p \le \varepsilon\\
		&\geq (1-\varepsilon)^3\cdot \frac{1}{\sigma} \cdot\sum_{i=1}^N (2\sigma p)^2 \cdot (1-2\sigma p)^{i-1}  \cdot c_i &&\mid \text{\cref{claim:auxiliary}}\\
		&= \ (1-\varepsilon)^3 \cdot \frac{1}{\sigma} \cdot\left( 2 \sigma \gamma + (1-2\sigma p)^{\frac{\gamma}{p}} - (1+2\sigma p N)\cdot (1- 2\sigma p)^{N} \right) &&\mid \text{\cref{lem:choice_of_p_and_n}~\ref{item:small_p_two}}\\
		&\ge \ (1-\varepsilon)^3 \cdot   \frac{1}{\sigma} \cdot\left( 2 \sigma \gamma + e^{-2\sigma\gamma} -\varepsilon - (1+2\sigma p N) \cdot(1- 2\sigma p)^{N} \right) &\\
		&\ge \ (1-\varepsilon)^3 \cdot   \frac{1}{\sigma} \cdot\left( 2 \sigma \gamma + e^{-2\sigma\gamma} -2\varepsilon \right)\\
		&\ge \ (1-\varepsilon)^3 \cdot (1 -2\varepsilon)\cdot   \frac{1}{\sigma} \cdot\left( 2 \sigma \gamma + e^{-2\sigma\gamma}\right).
	\end{align*}  
Note that for the penultimate inequality we used \cref{lem:choice_of_p_and_n}~\ref{item:large_N} for $x=2\sigma pN \ge \epsilon^2pn-3$ and 
\[(1+x)\cdot\left(1-\frac{x}{N}\right)^N \ \leq \ (1+x)\cdot e^{-x} \ \le \ (1+x)\cdot e^{-p\varepsilon^2x}.\]
The last inequality follows since 
$x+e^{-x}\geq 1$ for all $x\in\mathbb{R}$.
	
	If we compute a TSP tour on the sampled customers that costs $\alpha$ times more than optimal, 
	the computed master tour has expected cost at least $\alpha$ times \eqref{eq:master_tour_cost}.
Hence, adding up the expected cost of the master tour and the expected connection cost yields at least
	\[ (1-\epsilon)^3\cdot (1-4\epsilon) \cdot\left( \alpha(\sigma\gamma + e^{-\sigma\gamma}) + 2\gamma + \frac{1}{\sigma}e^{-2\sigma\gamma} \right) .\]
	\end{proof}
	\begin{proof}[Proof of \cref{thm:sampling_lb}] 
	Putting together \cref{prop:ub_opt} and \cref{prop:lb_sampling} and considering the limit $\varepsilon \to 0$,
	the ratio between the expected cost of the master route solution that we get from sampling
	and the expected cost of an optimum a priori tour is at least
	\begin{equation} \label{eq:lowerboundonsampling}
		\frac{\alpha(\sigma\gamma + e^{-\sigma \gamma}) + 2\gamma + \frac1\sigma e^{-2\sigma\gamma}}{\gamma + e^{-\gamma}}.
	\end{equation}
	For any fixed $\alpha$ and $\gamma$, this ratio is minimized for the unique positive $\sigma$ for which
	\[
	\sigma^2\alpha\gamma(e^{2\sigma\gamma}-e^{\sigma\gamma}) \ = \ 1+2\sigma\gamma,
	\]
	but apparently there is no closed-form solution. 
	Optimizing this term numerically gives the bounds shown in \cref{fig:lb_ex_sampling}.
	Note that due to uniform activation probabilities, the function $f$ in the sampling algorithm is irrelevant except for 
	the value of $f(p)$, which is completely determined by $\sigma$. 
\end{proof}

\section{Upper bound on the master route ratio}
\label{sec:mastrerouteratio_ub}

Our proof of the upper bound on the master route ratio (for normalized \textsc{a priori TSP} instances)
follows a similar line as the proof of \cref{thm:ub_sampling} in \cref{sec:sampling_ub}.
However, there are some important differences and further complications.
In particular, it is not easy to bound the expected cost for connecting the active customers to the master tour 
in terms of the buckets introduced in \cref{section:sampling_singleLP}. This will require another level of aggregation. As in \cref{sec:sampling_ub}, let $\beta, b_0$ be constants that we will choose later.

\subsection{An optimization problem for the master route ratio}

Consider a normalized instance, and let $p$ denote the (uniform) activation probability.
Let $T^*$ be a fixed optimum a priori tour, with customers appearing in the order $v_0,v_1,\ldots,v_{n-1}$; 
here $v_0$ denotes the depot. Let $v_i\coloneqq v_0$ for $i<0$ or $i>n-1$.
As in \cref{sec:sampling_OP}, we define for $k\in\mathbb{Z}_{\ge 1}$
\[
C_k \ \coloneqq \ p^2 \cdot \sum_{j\in\mathbb{Z}} c(v_j,v_{j+k}),
\]
which are nonnegative variables satisfying the triangle inequality 
\begin{equation}\label{eq:triangle_mrr}
	C_{i+j} \ \le \ C_i + C_j 
\end{equation}
for all $i,j \ge 1$.
As seen in \cref{prop:opt_C_i}, the expected cost of $T^*$ is exactly
\begin{equation} \label{eq:costofoptimumsolution}
	\sum_{i=1}^{\infty} (1-p)^{i-1}\cdot C_i .
\end{equation} 

We now design master route solutions.
For $k \ge 1$, consider a master route solution in which the master tour contains the depot and, 
with some offset $h\in\{1,\ldots,k\}$, every $k$-th customer on $T^*$, i.e., 
$\{v_j:j\in\mathbb{Z}, j\equiv h  \pmod k\}$.
Note that this means that for $h \ge n$, the master tour consists only of the depot (recall that $v_j = v_0$ for $j < 0$ and $j > n-1$).
We will now show that, for any fixed $k$, the expected cost of the best such solution is at most
\begin{equation} \label{eq:costofsampledmasterroutesolution}
	\frac{1}{kp^2}\left( C_k + 2 p\cdot \sum_{i=1}^{k-1} \min\left\{\big. C_i,\, C_{k-i} \right\} \right).
\end{equation}
If we choose the offset uniformly at random, the master tour has expected cost $\frac{C_k}{kp^2}$. 
Indeed, for $k < n$, this directly follows by the definition of $C_k$, and for $k \ge n$, the master tour has expected cost 
\begin{align*}
	\Prob[h < n] \cdot \frac{C_{n-1}}{(n-1)p^2} &\ = \ \frac{C_{n-1}}{kp^2} \ = \ \frac{C_k}{kp^2} 
\end{align*}	
as claimed.

Now we bound the expected cost of connecting the active customers to the master tour. 
Again we do this by considering only the following two options for each active customer $v$:
the first sampled customer that we encounter when traversing $T^*$ from $v$ in either direction.
Connecting to other sampled customers may be cheaper, but we ignore this again and still obtain an upper bound. 
Now, for offset $h$, we obtain an upper bound on the connection cost of
\[2 p\cdot \!\!\sum_{\substack{j\in \mathbb{Z}:\\j\equiv h \!\!\!\!\pmod k}}\sum_{i=1}^{k-1} \ \min\{c(v_{j},v_{j+i}),c(v_{j+i},v_{j+k})\}\]
since the master tour  contains precisely the customers $v_j$ with $j\equiv h\pmod k$. Choosing $h$ uniformly at random, this results in a bound on the connection cost of
\begin{align*}
	&\frac{2 p}{k}\cdot\sum_{i=1}^{k-1}\sum_{j\in\mathbb{Z}}\min\{c(v_j,v_{j+i}),c(v_{j+i},v_{j+k})\}\\
	&\leq \ \frac{2 p}{k}\cdot \sum_{i=1}^{k-1}\min\left\{\sum_{j\in\mathbb{Z}} c(v_j,v_{j+i}),\sum_{j\in\mathbb{Z}} c(v_j,v_{j+k-i})\right\}\\
	&= \ \frac{2}{p k}\cdot\sum_{i=1}^{k-1}\min\{C_i,C_{k-i}\} .
\end{align*} 
This concludes the proof that \eqref{eq:costofsampledmasterroutesolution} is an upper bound on the best ``equidistant'' master route solution,
and hence on the best master route solution per se.

So for every $k\in\mathbb{N}$, 
the ratio of \eqref{eq:costofsampledmasterroutesolution} to \eqref{eq:costofoptimumsolution} 
is an upper bound on the master route ratio for that instance.
In other words, minimizing \eqref{eq:costofoptimumsolution} subject to the constraints that
\eqref{eq:costofsampledmasterroutesolution} is equal to 1 and the $C_i$ are nonnegative and
satisfy the triangle inequality \eqref{eq:triangle_mrr} yields the reciprocal of an upper bound
on the master route ratio of all normalized instances with activation probability $p$.
Note that only requiring that \eqref{eq:costofsampledmasterroutesolution} is at least 1 does not change the minimum.
Again, the number $n$ of customers appears neither in \eqref{eq:costofoptimumsolution} nor in \eqref{eq:costofsampledmasterroutesolution}.
We arrive at the following optimization problem:

\begin{align}
	\min \sum_{i=1}^{\infty} (1-p)^{i-1}\cdot C_i  &&  \tag{Master-Route-Ratio-OP} \label{eq:exact_lp_for_masterrouteratio} \\
	\text{subject to } \hspace*{5cm} 
	C_i &\ \geq \ 0 &\text{for } i\in\mathbb{N} \label{eq:lp:Monotonicity}\\
	C_i+C_j &\ \geq \ C_{i+j}  &\text{for } i,j\in\mathbb{N} 
	\label{eq:lp:TriangleInequalities}\\
	C_k + 2p\cdot \sum_{i=1}^{k-1} \min\left\{\big. C_i,\, C_{k-i} \right\} &\ \geq \ kp^2 &\text{for } k \in \mathbb{N}  \label{eq:lp:MasterRoutes}.
\end{align}

We have proved:

\begin{lemma}
	Let $p>0$.
	The reciprocal of the value of \eqref{eq:exact_lp_for_masterrouteratio}
	is an upper bound on the master route ratio for $p$-normalized instances. 
	\hfill\qed
\end{lemma}

\subsection{Obtaining a single linear program}

The optimization problem \eqref{eq:exact_lp_for_masterrouteratio} has infinitely many variables, 
but one could omit the terms for, say, $i>\frac{100}{p}$ in the infinite sum while still getting a good upper bound. 
One could also omit the constraints of type \eqref{eq:lp:MasterRoutes} for, say, $k > \frac{100}{p}$ 
while still getting a good upper bound. 

However, we would still have an infinite set of LPs (one for each choice of $p$), and, by \cref{lemma:wlog_normalized}, 
we should consider the limit for $p\to 0$.

In the following, exactly as in \cref{section:sampling_singleLP}, we require that $p$ is of the form $p = \frac{\beta}{b}$ for some odd integer $b \ge b_0$. Note that $p \to 0$ as $b \to \infty$. In order to obtain a single optimization problem for all such values of $p$,
we again put subsequent $C_i$'s into buckets of size $b$: We define
\begin{equation} \label{eq:mrr_define_buckets}
	B_i \ \coloneqq \ \sum\limits_{j=\max\{1,ib-\frac{b-1}2\}}^{ib+\frac{b-1}2} C_{j}
\end{equation}
for $0 \le i \le N$ for some integer $N$ that we will choose later. In the following, we show that we can use the constraints in \eqref{eq:exact_lp_for_masterrouteratio} to generate finitely many (slightly relaxed) constraints that only depend on these buckets. 
Note that for all $i,j \in \mathbb{Z}_{\ge 1}$ with $i+j \le N$ we have $B_{i+j} \le B_i + B_j$ by \cref{prop:triangle_buckets}.

For each $1 \le i \le N$, we sum the constraints of type \eqref{eq:lp:MasterRoutes} for 
all $k$ in the $i$-th bucket (i.e., $k= ib-\frac{b-1}2, \dots, ib + \frac{b-1}2$), yielding
\begin{align}\label{eq:new_mr_constraints}
	B_i + 2p \sum_{k= -\frac{b-1}2}^{\frac{b-1}2} \sum_{j=1}^{ib+k-1} \min\left\{\big. C_j,\, C_{ib+k-j} \right\} 
	\ \ge \ \sum_{k= -\frac{b-1}2}^{\frac{b-1}2} (ib + k) p^2 \ = \  i (bp)^2 . 
\end{align}
The left-hand side of \eqref{eq:new_mr_constraints} still contains the $C_i$ variables, and we want to replace them
by the new $B_i$ variables. 

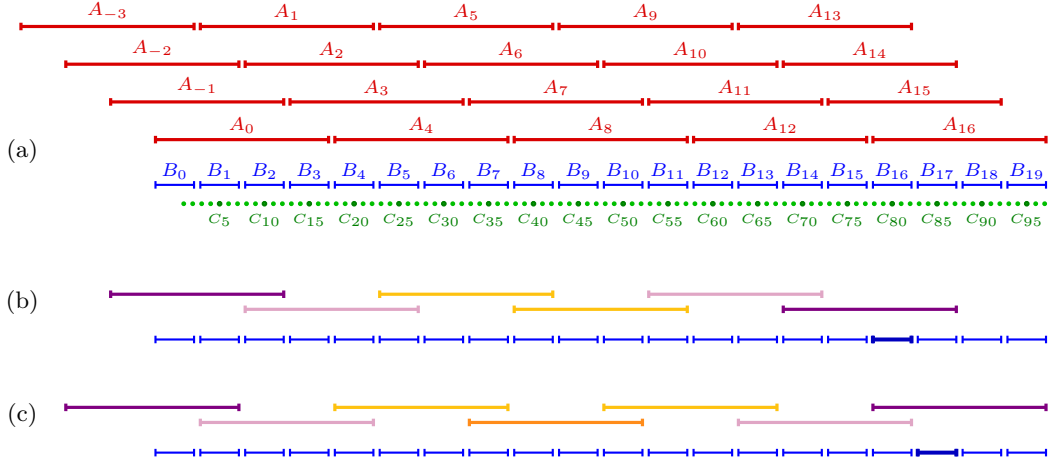
\begin{figure}[ht]
	\begin{center}
		\begin{tikzpicture}[xscale=0.59, thick]
			\tikzstyle{singlevertex}=[darkgreen,circle,draw,minimum size=1,inner sep=0pt]
			\tikzstyle{mainsinglevertex}=[verydarkgreen,circle,draw,minimum size=1.3,inner sep=0pt]
			\tikzstyle{bucket}=[blue]
			\tikzstyle{mainbucket}=[darkblue, line width=1.5]
			\tikzstyle{interval}=[red!85!black, line width=1.2]
			\begin{scope}[shift={(0,0)}]
				\node at (-3,0.7) {\small (a)};
				\foreach [evaluate=\i as \pos using (\i / 5)] \i in {3, 4, ..., 99} 
				{
					\node[singlevertex] at (\pos,0) {};
				}
				\foreach 
				[evaluate=\i as \pos using (\i / 5),
				evaluate=\i as \index using (int(\i -2))] 
				\i in {7, 12, ..., 97} 
				{
					\node[mainsinglevertex] at (\pos,0) {};
					\node[verydarkgreen,below] at (\pos, 0) {\tiny$C_{\index}$};
				}
				\foreach 
				[evaluate=\i as \end using (\i + 4/5 + 0.03), 
				evaluate=\i as \begin using (\i - 0.03)] 
				\i in {0, 1, ..., 19} 
				{
					\draw[bucket] (\begin,0.25) to node[above=-0.5mm]{\scriptsize $B_{\i}$} (\end,0.25);
					\draw[bucket] (\begin,0.2) -- (\begin,0.3);
					\draw[bucket] (\end,0.2) -- (\end,0.3);
				}
				\foreach 
				[evaluate=\offset as \ycoor using (0.85+0.5*\offset),
				evaluate=\offset as \ya using (0.8+0.5*\offset),
				evaluate=\offset as \yb using (0.9+0.5*\offset)] 
				\offset in {0, 1, 2, 3} 
				{
					\foreach 
					[evaluate=\i as \end using (\i - \offset + 3 + 4/5 + 0.03), 
					evaluate=\i as \begin using (\i - \offset - 0.03),
					evaluate=\i as \index using int(\i - \offset)] 
					\i in {0, 4, ..., 16} 
					{
						\draw[interval] (\begin,\ycoor) to node[above=-0.5mm]{\scriptsize $A_{\index}$} (\end,\ycoor);
						\draw[interval] (\begin,\ya) -- (\begin,\yb);
						\draw[interval] (\end,\ya) -- (\end,\yb);
					}
				}
			\end{scope}
			
			\begin{scope}[shift={(0,-2)}]
				\node at (-3,0.7) {\small (b)};
				\foreach 
				[evaluate=\i as \end using (\i + 4/5 + 0.03), 
				evaluate=\i as \begin using (\i - 0.03)] 
				\i in {0, 1, ..., 19} 
				{
					\draw[bucket] (\begin,0.2) -- (\end,0.2);
					\draw[bucket] (\begin,0.15) -- (\begin,0.25);
					\draw[bucket] (\end,0.15) -- (\end,0.25);
				}
				\foreach 
				[evaluate=\i as \end using (\i + 4/5 + 0.03), 
				evaluate=\i as \begin using (\i - 0.03)] 
				\i in {16} 
				{
					\draw[mainbucket] (\begin,0.2) -- (\end,0.2);
					\draw[mainbucket] (\begin,0.15) -- (\begin,0.25);
					\draw[mainbucket] (\end,0.15) -- (\end,0.25);
				}

				\foreach 
				\offset in {1} 
				{
					\foreach 
					[evaluate=\i as \end using (\i - \offset + 3 + 4/5 + 0.03), 
					evaluate=\i as \begin using (\i - \offset - 0.03),
					evaluate=\y as \ycoor using (0.4+0.2*\y),
					evaluate=\y as \ya using (0.35+0.2*\y),
					evaluate=\y as \yb using (0.45+0.2*\y)]  
					\i/\y/\c in {0/2/{blue!50!red}, 3/1/{blue!35!red!35!white}, 6/2/{yellow!50!orange}, 9/1/{yellow!50!orange}, 12/2/{blue!35!red!35!white}, 15/1/{blue!50!red}} 
					{
						\draw[interval,\c] (\begin,\ycoor) -- (\end,\ycoor);
						\draw[interval,\c] (\begin,\ya) -- (\begin,\yb);
						\draw[interval,\c] (\end,\ya) -- (\end,\yb);
					}
				}
				
			\end{scope}
			
			\begin{scope}[shift={(0,-3.5)}]
				\node at (-3,0.7) {\small (c)};
				\foreach 
				[evaluate=\i as \end using (\i + 4/5 + 0.03), 
				evaluate=\i as \begin using (\i - 0.03)] 
				\i in {0, 1, ..., 19} 
				{
					\draw[bucket] (\begin,0.2) -- (\end,0.2);
					\draw[bucket] (\begin,0.15) -- (\begin,0.25);
					\draw[bucket] (\end,0.15) -- (\end,0.25);
				}
				\foreach 
				[evaluate=\i as \end using (\i + 4/5 + 0.03), 
				evaluate=\i as \begin using (\i - 0.03)] 
				\i in {17} 
				{
					\draw[mainbucket] (\begin,0.2) -- (\end,0.2);
					\draw[mainbucket] (\begin,0.15) -- (\begin,0.25);
					\draw[mainbucket] (\end,0.15) -- (\end,0.25);
				}

				\foreach 
				\offset in {2} 
				{
					\foreach 
					[evaluate=\i as \end using (\i - \offset + 3 + 4/5 + 0.03), 
					evaluate=\i as \begin using (\i - \offset - 0.03),
					evaluate=\y as \ycoor using (0.4+0.2*\y),
					evaluate=\y as \ya using (0.35+0.2*\y),
					evaluate=\y as \yb using (0.45+0.2*\y)]  
					\i/\y/\c in {0/2/{blue!50!red}, 3/1/{blue!35!red!35!white}, 6/2/{yellow!50!orange}, 9/1/{orange!90!white}, 12/2/{yellow!50!orange}, 15/1/{blue!35!red!35!white}, 18/2/{blue!50!red}} 
					{
						\draw[interval,\c] (\begin,\ycoor) -- (\end,\ycoor);
						\draw[interval,\c] (\begin,\ya) -- (\begin,\yb);
						\draw[interval,\c] (\end,\ya) -- (\end,\yb);
					}
				}
				
			\end{scope}
			
		\end{tikzpicture}
	\end{center}
	\caption{(a): The green dots stand for $C_1,C_2,\ldots$, and the centers of the buckets ($C_{ib}$ for $i\ge 1$) are highlighted.
		Here the bucket size is $b=5$, and the blue intervals show the buckets $B_0, B_1,B_2,\ldots$
		We now form bucket intervals, shown in red on the top (each consisting of $a+1=4$ buckets; here $a=3$; 
		we can think of adding empty buckets on the left).
		(b): Aggregating the constraints \eqref{eq:lp:MasterRoutes} for bucket $B_{16}$ (i.e., $k\in\{78,79,80,81,82\}$)
		yields \eqref{eq:new_mr_constraints} and then \eqref{eq:final_mr_constraints} for $i=16$; 
		here we collect the terms on the left-hand side according to the bucket intervals shown.
		(c): The same for bucket $B_{17}$.
		\label{fig:bucket_intervals}}
\end{figure}

Since it is not possible to express the minima in the left-hand side of \eqref{eq:new_mr_constraints} in terms of the buckets without an additional loss,
we will need to be a bit pessimistic here.
In order to not lose too much, we form bucket intervals
\begin{equation}
	\label{eq:bucket_interval}
	A_j \ \coloneqq \ \sum_{\ell=\max\{j,0\}}^{\min\{j+a,N\}} B_{\ell}
\end{equation}
for $j=-a,\ldots,N$,
each consisting of $a+1$ subsequent buckets (except for $j<0$ or $j > N-a$), where $a$ is some odd integer (think of $a$ large but $abp$ small).
Roughly speaking, the purpose of the bucket intervals is the following: When deciding to which sampled neighbor on $T^*$ we connect the active customers, we have to make the same decision for all customers in the same bucket interval.
It turns out that this way we will lose only a fraction proportional to $\frac{1}{a}$.
See \cref{fig:bucket_intervals} for an illustration.

Then, choosing some offset $h\in\{0,\dots, a-1\}$, 
we bound the double sum on the left-hand side of \eqref{eq:new_mr_constraints} as follows, where we use $C_j = 0$ for $j \le 0$ and $j > Nb + \frac{b-1}2$ (i.e., $C_j$ does not occur in any bucket), and $B_j = 0$ for $j<0$ and $j > N$ in intermediate steps to simplify the terms.
\begin{align*}
	&	\sum_{k= -\frac{b-1}2}^{\frac{b-1}2} \sum_{j=1}^{ib+k-1} \min\left\{\big. C_j,\, C_{ib+k-j} \right\} \\
	&\ \le \ \sum_{k= -\frac{b-1}2}^{\frac{b-1}2} \, \sum_{j=1}^{ib+k-1+hb} \min\left\{\big. C_{j-hb},\, C_{ib+k-j+hb} \right\} \\
	& \ \le \ \sum_{k= -\frac{b-1}2}^{\frac{b-1}2} \sum_{j=1}^{\left\lceil \frac{h+i+1}a \right\rceil} \sum_{\ell = 1}^{ab}\min\{C_{(j-1)ab+\ell-hb},\, C_{ib+k-(j-1)ab-\ell+hb} \} \\
	& \ \le \ \sum_{k= -\frac{b-1}2}^{\frac{b-1}2} \sum_{j=1}^{\left\lceil \frac{h+i+1}a \right\rceil} \min\left\{\sum_{\ell = 1}^{ab}C_{(j-1)ab+\ell-hb},\,\sum_{\ell = 1}^{ab} C_{ib+k-(j-1)ab-\ell+hb} \right\} \\
	& \ = \ \sum_{k= -\frac{b-1}2}^{\frac{b-1}2} \sum_{j=1}^{\left\lceil \frac{h+i+1}a \right\rceil} \min\left\{\big. \sum_{\ell = 1}^{ab} C_{(j-1)ab+\ell-hb},\, \sum_{\ell=1}^{ab} C_{ib+k-jab+\ell-1+hb} \right\} \\
	& \ \le \ \sum_{k= -\frac{b-1}2}^{\frac{b-1}2} \sum_{j=1}^{\left\lceil \frac{h+i+1}a \right\rceil} \min\left\{\big. \sum_{\ell = 0}^{a} B_{(j-1)a+\ell-h},\,
	\sum_{\ell= 0}^a B_{i-ja+\ell+h} \right\} \\
	& \ = \ b \sum_{j=1}^{\left\lceil \frac{h+i+1}a \right\rceil} \min\left\{\big. A_{(j-1)a-h},\, A_{i-ja+h} \right\} . 
\end{align*}
To minimize the loss, we choose $h_i = (i \mod a)$ because this implies that different buckets are counted twice for different values of $i$, i.e., the buckets $B_{aj-h_i}$ ($j\ge 1$) are counted twice,
as \cref{fig:bucket_intervals}~(b) and (c) indicate, but note that this consideration is not needed for correctness. 
In any case, we conclude that for any $1 \le i \le N$,
\begin{equation}
	\label{eq:final_mr_constraints}
	B_i + 2bp \sum_{j=1}^{\left\lceil \frac{h_i+i+1}a \right\rceil} \min\left\{\big. A_{(j-1)a-h_i},\, A_{i-ja+h_i} \right\}
	\ \ge \ i(bp)^2
\end{equation}
is a relaxation of \eqref{eq:new_mr_constraints}, 
and hence of \eqref{eq:lp:MasterRoutes} summed for $k=ib-\frac{b-1}{2},\ldots,ib+\frac{b-1}{2}$.

Therefore, we obtain a lower bound on the value of \eqref{eq:exact_lp_for_masterrouteratio} by 
minimizing \mbox{$\sum_{i=1}^{\infty} (1-p)^{i-1}\cdot C_i$} subject to 
\eqref{eq:bucket_interval} and \eqref{eq:final_mr_constraints} and
$B_{i+j}\ge B_i+B_j$ for $i,j\ge 1$ with $i+j \le N$, and $B_i\ge 0$ for $i\ge 0$. 

The objective function still contains infinitely many variables and depends on $p$, and we resolve this in the same way 
as in \cref{section:sampling_singleLP}. Again, as in \eqref{eq:lb_obj},
\begin{align}\label{eq:lb_obj_mrr}
	\sum_{i=1}^{\infty} (1-p)^{i-1}\cdot C_i 
	\ \ge \ \sum_{i=0}^{\infty}  (1-p)^{bi+\frac{b-1}2-1}\cdot B_i 
	\ \ge \ \sum_{i=0}^{N} (1-p)^{(i+\frac12)b}\cdot B_i.
\end{align}
Moreover, $p=\frac{\beta}{b}$ and $\lim_{b\to\infty}(1-\frac{\beta}{b})^{(i+\frac12) b} = e^{-(i+\frac12)\beta}$ for all $i=0,\ldots,N$.
Again, we will use this to replace the objective function by  $\sum_{i=0}^{N} e^{-(i+\frac12)\beta} \cdot B_i$ and still get an upper bound.

Finally, we omit many triangle inequalities and keep only $i\cdot B_1\ge B_i$ for $i=2,\ldots,N$. 
Moreover, we further introduce auxiliary variables for the minima in \eqref{eq:final_mr_constraints}.
Here is our final linear program:
\begin{align}
	\min \sum_{i=0}^{N} e^{-(i+\frac12)\beta} \cdot B_i &&  \tag{Master-Route-Ratio-LP}\label{eq:final_masterroute_LP}  \\
	\text{subject to } \hspace*{1.3cm} 
	i \cdot B_1 &\ \geq \ B_i  &\text{for } i=2, \ldots, N \label{eq:lp:Sublinear} \\
	B_i +  2\beta  \sum_{j=1}^{\left\lceil \frac{h_i+i+1}a \right\rceil} M_{j,i}
	&\ \geq \ i\beta^2 &\text{for } i=1,\ldots,N \label{eq:lp:mr_final} \\
	A_{(j-1)a-h_i} &\ \geq \ M_{j,i} &\text{for } i=1,\ldots,N \text{ and } j=1,\ldots, \textstyle{\left\lceil \frac{h_i+i+1}a \right\rceil} \label{eq:lp:min1}\\
	A_{i-ja+h_i} &\ \geq \ M_{j,i}&\text{for } i=1,\ldots,N \text{ and } j=1,\ldots, \textstyle{\left\lceil \frac{h_i+i+1}a \right\rceil} \label{eq:lp:min2}\\
	\sum_{\ell=\max\{i,\, 0\}}^{\min\{i+a,\, N\}} B_{\ell} &\ = \ A_i &\text{for } i = -a,\dots,N \label{eq:lp:larger_buckets}\\
	B,M &\ \ge \ 0 .
\end{align}

We conclude:

\begin{lemma}
	Let $N,a$ be integers with $a$ odd and let $\beta > 0$. Let $h_i \in \{0,\ldots,a-1\}$ for $i=1,\ldots,N$.
	The reciprocal of the optimum value of \eqref{eq:final_masterroute_LP}
	is an upper bound on the master route ratio for \textsc{a priori TSP} instances with depot.
\end{lemma}

\begin{proof}
	We proceed analogously to the proof of \cref{lem:primal_lp_sampling}.
	We compare the value of \eqref{eq:final_masterroute_LP} to the value of \eqref{eq:exact_lp_for_masterrouteratio}.
	We showed that for any feasible solution $C$ to \eqref{eq:exact_lp_for_masterrouteratio} we obtain a feasible solution
	$(A,B,M)$ to \eqref{eq:final_masterroute_LP} via \eqref{eq:mrr_define_buckets} and \eqref{eq:bucket_interval}
	and $M_{j,i}=\min\{A_{(j-1)a-h_i},A_{i-ja+h_i}\}$.
	
	Fix $\delta>0$. Then there exists $b_0\in\mathbb{N}$ such that for all odd integers $b\ge b_0$
	\begin{equation*}
		\sum_{i=0}^{N} e^{-(i+\frac12)\beta} \cdot B_i
		\ \le \ (1+\delta)\cdot\sum_{i=0}^{N}\left(1-\textstyle{\frac{\beta}{b}}\right)^{(i+\frac12)b} \cdot B_i
		\ \le \ (1+\delta)\cdot\sum_{i=0}^{\infty}\left(1-\textstyle{\frac{\beta}{b}}\right)^{i-1} \cdot C_i.
	\end{equation*} 
	Thus we conclude that the value of \eqref{eq:final_masterroute_LP} is at most $(1+\delta)$ times the value of \eqref{eq:exact_lp_for_masterrouteratio}
	with $p=\frac{\beta}{b}$ for all odd integers $b\ge b_0$.
	Hence, by \cref{lemma:sampling-OP},
	$(1+\delta)$ times the reciprocal of the value of \eqref{eq:final_masterroute_LP} is an upper bound on the master route ratio 
	for all $\frac{\beta}{b}$-normalized instances for all odd integers $b\ge b_0$. 
	By \cref{lemma:wlog_normalized}, the same bound then holds for all instances with depot.
	Since this bound holds for all $\delta>0$, it also holds for $\delta=0$.
\end{proof}

\subsection{Solving the dual LP}

Now we dualize the LP~\eqref{eq:final_masterroute_LP}. For this, we introduce variables $(x_i)_{i=2,\ldots,N}$ for the inequalities of type \eqref{eq:lp:Sublinear}, variables $(y_i)_{i=1, \ldots, N}$ for the inequalities of type \eqref{eq:lp:mr_final}, variables $v_{i,j}$ and $w_{i,j}$ for the inequalities of type \eqref{eq:lp:min1} and \eqref{eq:lp:min2}, respectively, and variables $(z_i)_{i=-a,\ldots,N}$ for the inequalities of type \eqref{eq:lp:larger_buckets}.

{\small
	\begin{align}
		\max \sum_{i=1}^N i \beta^2 \cdot y_i  & & \tag{Dual-Master-Route-Ratio-LP}\label{eq:dual_of_masterroute_LP} \\
		\text{subject to } \quad y_i + \sum_{j=i-a}^{i} z_j + \mathds{1}_{i=1} \sum_{j = 2}^N j \cdot x_j - \mathds{1}_{2 \le i \le N} \cdot x_i&\ \le  \ 	e^{-(i+\frac12)\beta} 
		&\text{for } i=0,\dots,N \notag \\
		2\beta \cdot y_{i}  &\ \le \ v_{j,i} + w_{j,i} &\hspace*{-1cm}\text{for } i= 1,\dots, N \notag \\
		&&\hspace*{-1cm}\text{and } j=1,\dots, \left\lceil \textstyle{\frac{h_i+i+1}a} \right\rceil \notag \\ 
		\sum_{i=1}^N \sum_{j=1}^{\left\lceil\frac{h_i+i+1}{a}\right\rceil} \!
		\left( \mathds{1}_{k=(j-1)a-h_i} \cdot v_{j,i} + \mathds{1}_{k=i-ja+h_i} \cdot w_{j,i} \right)& \ = \ z_k
		&\text{for } k = -a,\dots,N \notag \\
		x,y,v,w &\ \ge \ 0 . \notag
	\end{align}
}

\begin{corollary}
	Let $N,a$ be integers with $a$ odd and let $\beta > 0$. Let $h_i \in \{0,\ldots,a-1\}$ for $i=1,\ldots,N$.
	For any feasible solution $(x,y,v,w,z)$ to \eqref{eq:dual_of_masterroute_LP}, $\left(\sum_{i=1}^N i \beta^2 \cdot y_i\right)^{-1}$ 
	is an upper bound on the master route ratio restricted to \textsc{a priori TSP} instances with depot. 
	\hfill\qed
\end{corollary}	

We have computed a dual solution using Gurobi 10.0.1 with $\beta = \frac1{400}$, $a = 199$, $N = 4000$, and $h_i = (i \mod a)$ for $i=1,\ldots,N$, 
yielding an upper bound of $2.584$ and thus proving \cref{thm:mrr}. 
The dual solution and a Python script that verifies that this is indeed a feasible solution to \eqref{eq:dual_of_masterroute_LP} can be found at \url{https://doi.org/10.60507/FK2/JCUIRI}.

\section{Lower bound on the master route ratio} \label{subsection:lowerboundonmasterrouteratio}

We now present an example that proves that the master route ratio is larger than $2.541$.

\begin{theorem} \label{thm:lowerboundonmasterrouteratio}
	The master route ratio is at least $\frac 1{1-e^{-\frac12}} > 2.541$.
\end{theorem}

\begin{proof}
	We provide a sequence of \textsc{a priori TSP} instances where the
	master route ratio converges to $\frac1{1-e^{-\frac12}}$. 
	Consider the complete graph on $n+1$ vertices $d,v_1,\dots,v_n$,
	and let the distance between any two vertices be 1. The vertex $d$ is our depot,
	and each other vertex is replaced by a group of $m$ customers that have distance 0 from each other. 
	More formally, we have
	$$V = \{d\}\cup\bigcup\limits_{i = 1}^nV_i, \qquad V_i = \{v_{i,1},\dots, v_{i,m}\}$$
	and a distance function given by $c(d, v_{i,j}) = 1$ and
	$$c(v_{i, j}, v_{i',j'}) \ = \ \begin{cases}
		1 & i \neq i'\\
		0 & i = i'\\
	\end{cases}.$$
	All activation probabilities are given by $p(v_{i,j}) = \frac{1}{2m}$, except for $p(d)=1$.
	
	Consider a master route solution for this instance.
	We may assume that the depot $d$ belongs to the master tour because otherwise 
	including it can only make the master route solution better. 
	Suppose the master tour visits $k$ of the groups, for some $k\in\{0,\ldots,n\}$.
	The cost of the master tour is then $0$ if $k=0$ and $k+1$ otherwise.
	Connecting customers to the master tour does not cost anything if there is a customer in the same group that belongs to the master tour. 
	Each of the other $m(n-k)$ customers has to be connected to the master tour with probability $\frac{1}{2m}$ and cost $2\cdot 1$,
	yielding a total expected connection cost of $n-k$. 
	Hence the expected cost of this master route solution is
	$$\begin{cases}
		n &\text{if } k = 0 \\
		k+1 + (n-k) & \text{if }k\ge 1
	\end{cases}.$$
	Thus in the best master route solution, the master tour consists only of the depot ($k=0$), and the expected cost is $n$.
	
	The optimum a priori tour visits the customers in each group consecutively. 
	Its cost only depends on the number of groups in which at least one customer is active.
	The probability that a group $V_i$ contains an active customer is $q\coloneqq1-(1-\frac{1}{2m})^m$, which
	converges to $1-e^{-\frac{1}{2}}$ as $m\to\infty$.  
	For every fixed $n$ there is thus a sufficiently large $m$ such that the expected number of active groups is at most 
	$$nq \ \leq \ n\left(1-e^{-\frac 12}\right) + 1.$$
	The cost of the resulting tour visiting the active customers is at most the number of active groups plus one (for the depot $d$).
	Hence the expected cost of the optimum a priori tour is at most
	$$nq + 1 \ \leq \ n\left(1-e^{-\frac12}\right) + 2 .$$ 
	This implies that the master route ratio is at least
	$$\frac{n}{n\left(1-e^{-\frac12}\right) + 2} \ \xrightarrow{n\to \infty} \ \frac 1{1-e^{-\frac12}}.$$
\end{proof}

\section{Introducing a depot (Proof of \cref{thm:can_assume_depot})}
\label{section:wlog_depot}

In this section, we show that assuming that the instance contains a \emph{depot}, 
i.e., an element $d\in V$ with $p(d)=1$, is no severe restriction.
We remark that although we can condition on the event that at least two customers are active
(because otherwise the cost is zero regardless of the a priori tour),
we cannot simply guess these a priori and declare one of them to be the depot.
As mentioned earlier, the previous works \cite{ShmT08} and \cite{Zuy11} provided ad hoc proofs that \emph{their} 
algorithms generalize to the non-depot case.
\cref{thm:can_assume_depot} yields a general reduction, losing only an arbitrarily small constant.

The proof consists of two parts. 
First we show that for instances whose total activation probability is bounded by a constant, we can find
an almost optimal master route solution, yielding a $(3+\varepsilon)$-approximation. 
The second part is to prove that otherwise it does not harm much to declare one customer to be the depot.

For instances in which the expected number of active customers is small (bounded by a constant)
it is crucial to take into account that the cost of an a priori tour cut short to fewer than two customers is zero:
For example, if $c(v,w)=1$ for all $v,w\in V$ with $v\not=w$ and $p(v)=\frac{\epsilon}{n}$ for all $v\in V$
(where $n=|V|$ and $\epsilon>0$ tends to zero), then $\Opt\approx \epsilon^2$ 
but the standard analysis of any master route solution (even if consisting only of a single point) 
yields cost at least roughly $2\epsilon$. However, using the fact that the cost is zero if fewer than two customers are active, Shmoys and Talwar~\cite{ShmT08} show that the a priori tour resulting from 
an optimum master route solution is \emph{always} at most a factor 3 worse than an optimum a priori tour.
The other good news is that a near-optimal master route solution can be found in polynomial time
if the expected number of active customers is bounded by a constant, which has some similarities to a PTAS found by 
Eisenbrand, Grandoni, Rothvo{\ss} and Sch\"afer~\cite{EGRS10} for connected facility location in a bounded case.
We need the following fact on the sum of independent Bernoulli random variables:

\begin{lemma} \label{lemma:large_sample_unlikely}
	Let $\epsilon>0$ and $k>0$ be constants. Then there is a constant $N=N(k,\epsilon)$ such that the following holds.
	Let $X_1,\ldots,X_n$ be independent Bernoulli variables and $X=\sum_{i=1}^n X_i$ with $\Prob[X\ge 2]>0$ and $\Exp[X] \le k$.
	Then $\Prob[X > N \mid X\geq 2]\leq \epsilon$.
\end{lemma}

\begin{proof}
	Let $p_i:=\Prob[X_i=1]$ for $i=1,\ldots,n$. We have $\sum_{i=1}^n p_i =\Exp[X]\leq k$.
	
	Let $S:=\{i\in\{1,\ldots,n\}: p_i \le \frac{1}{2}\}$ be the indices of the Bernoulli variables with small success probability.
	Let $X'=\sum_{i\in S} X_i$. We always have 
	\[X-X' \ \le \ \left|\left\{i\in\{1,\ldots,n\}: p_i>\frac{1}{2}\right\}\right| \ < \ 2k.\]
	To bound the probability that $X'$ is large, we compute, for $m>2$,
	\begin{align*}
		\Prob[X'=m] &\ = \ \sum_{T\in {S\choose m}} \,\prod_{j\in T} p_j \cdot \prod_{j\in S\setminus T}(1-p_j) \\
		&\ = \ \frac{1}{m} \sum_{i\in S} p_i \cdot \sum_{T\in {S\setminus\{i\} \choose m-1}} \,\prod_{j\in T}p_j \cdot \prod_{j\in S\setminus( T\cup \{i\})}(1-p_j) \\
		&\ = \ \frac{1}{m} \sum_{i\in S} \frac{p_i}{1-p_i}\cdot\sum_{T\in {S\setminus\{i\} \choose m-1}}\prod_{j\in T}p_j \cdot \prod_{j\in S\setminus T}(1-p_j)\\
		&\ \le \ \frac{1}{m} \sum_{i\in S} \frac{p_i}{1-p_i}\cdot\sum_{T\in {S\choose m-1}}\prod_{j\in T}p_j \cdot \prod_{j\in S\setminus T}(1-p_j)\\
		&\ = \ \frac{1}{m} \sum_{i\in S} \frac{p_i}{1-p_i} \cdot \Prob[X'=m-1] \\
		&\ \le \ \frac{1}{m} \sum_{i\in S} 2p_i \cdot \Prob[X'=m-1] \\
		&\ \le \ \frac{2k}{m} \cdot \Prob[X'=m-1],
	\end{align*}
	where we used $p_i\le\frac{1}{2}$ for $i\in S$ in the second inequality.
	An iterative application yields 
	\[\Prob[X' = m] \ \leq \ \frac{2\cdot (2k)^{m-2}}{m!}\cdot \Prob[X'=2].\]
	Hence, for any integer $\ell \ge 2ek$, 
	\begin{align*}
		\Prob[X \ge 2k+\ell] &\ \le \ \Prob[X'\ge \ell+1] \ \le \ \sum_{m=\ell+1}^{\infty}\frac{2\cdot (2k)^{m-2}}{m!}\cdot \Prob[X'=2]\\
		& \ \le \ \frac{2}{\ell}\cdot \Prob[X'= 2] \ \le \ \frac{2}{\ell}\cdot \Prob[X\ge 2],
	\end{align*}
	where the third inequality follows (with Stirling's formula) from 
	\begin{align*}
		\sum_{m=\ell+1}^{\infty}\frac{(2k)^{m-2}}{m!} &\ \le \ \sum_{m=\ell+1}^{\infty}\frac{(\frac{\ell}{e})^{m-2}}{m!} 
		\ \le \ \frac{e(\frac{\ell}{e})^\ell}{\ell(\ell+1) \ell!} \cdot \sum_{i=0}^{\infty}e^{-i}\\
		&\ = \ \frac{e^2}{(e-1)\ell(\ell+1)} \cdot \frac{(\frac{\ell}{e})^\ell}{\ell!} \ \le \ \frac{e^2}{\sqrt{2\pi \ell}(e-1)\ell(\ell+1)}
		\ \le \ \frac{1}{\ell}\end{align*}
	for $\ell\ge 1$.
	Setting $\ell=\left\lceil\max\{2ek,\frac{2}{\epsilon}\}\right\rceil$ and $N=2k+\ell$ finishes the proof.
\end{proof}

Now we are ready to prove the following: 

\begin{lemma} \label{lemma_low_activity}
	Let $k>0$ and $\epsilon>0$ be constants. 
	Then we can find an a priori tour with expected cost at most $(3+\epsilon)\cdot\Opt$ for any given 
	\textsc{a priori TSP} instance $(V,c,p)$ with $\sum_{v\in V}p(v)\le k$ in polynomial time.
\end{lemma}

\begin{proof}
	Shmoys and Talwar~\cite{ShmT08} showed that if we randomly sample a subset $S\subseteq V$
	with at least two elements, where $S$ is chosen with probability 
	$(\Prob_{A\sim p}[A=S]) / (\Prob_{A\sim p}[|A|\ge 2])$,
	then any a priori tour $T_S$ corresponding to the master route solution with an optimum TSP tour for $S$ as master tour has expected cost 
	at most $3\cdot\Opt$.
	
	Now we exploit that $\sum_{v\in V}p(v)\le k$. We may assume $\epsilon \le 1$. 
	By Lemma~\ref{lemma:large_sample_unlikely} there is a constant $N=N(k,\frac{\epsilon}{4})$ such that,
	for the randomly chosen $S$, the probability that $|S|> N$ is at most $\frac{\epsilon}{4}$.
	Hence for at least one set $S$ with $2\le |S|\le N$ we have
	\begin{equation*}
		\Exp_{A\sim p}\left[c(T_S[A])\right] \ \le \ \frac{3}{1- \frac{\varepsilon}{4}} \cdot \Opt \ \le \  (3+\epsilon)\cdot\Opt.
	\end{equation*}
	
	Since $N$ is a constant, we can find such a set $S$ and an optimum TSP tour for $S$ by complete enumeration.
\end{proof}

By \cref{lemma_low_activity} we can assume that the expected activity is high. Then assuming a depot incurs an arbitrarily small loss only:

\begin{lemma} \label{lemma:depot_when_highactivity}
Let $\epsilon > 0$ and $\rho \geq 3$ be constants. If there is a (randomized) polynomial-time $\rho$-approximation algorithm for instances of the \textsc{a priori TSP} that have a depot $d$ with $p(d) = 1$, then there is a (randomized) polynomial-time $(\rho + \epsilon)$-approximation algorithm for instances $(V,c,p)$ with  $\sum_{v\in V} p(v) \ge \frac{2\rho}{\epsilon}$.
\end{lemma}

\begin{proof}
Let $p(V)\coloneqq\sum_{v\in V} p(v)$ and assume $p(V)\ge \frac{2\rho}{\epsilon}$. 
We try all $v\in V$, redefine $p(v)=1$ (i.e., make $v$ the depot), and call the $\rho$-approximation
	algorithm on the resulting instance. 
	
More precisely, for $v\in V$, let $p^{v=d}$ be defined by $p^{v=d}(u)=p(u)$ for $u\in V\setminus\{v\}$ and $p^{v=d}(v)=1$.
	In addition, let $\Opt^{v=d}$ denote the expected cost of an optimum a priori tour for the modified instance where we replace $p$ by $p^{v=d}$. 
	Calling the $\rho$-approximation for instances with depot yields a solution $T^v$ with expected cost
	$\Exp_{A\sim p^{v=d}}[c(T^v[A])]\leq \rho\cdot \Opt^{v=d}$.
	
	Note that for every TSP tour $T$ for $V$ and any $v\in V$, we have 
	\begin{align*}
		&\sum_{U\subseteq V} \Prob_{A\sim p}[A=U]\cdot c(T[U\cup \{v\}])\\
		& \ = \sum_{U\subseteq V: v\in U} \left(\big. \Prob_{A\sim p}[A=U\setminus\{v\}]+\Prob_{A\sim p}[A=U] \right)\cdot c(T[U])\\
		& \ = \sum_{U\subseteq V: v\in U} \ \prod_{u\in U\setminus\{v\}}p(u) \cdot \!\!\prod_{w\in V\setminus U}\!\! (1-p(w)) \cdot c(T[U])\\
		& \ = \ \sum_{U\subseteq V: v\in U} \Prob_{A\sim p^{v=d}}[A=U] \cdot c(T[U]) \\
		& \ = \ \Exp_{A\sim p^{v=d}}[c(T[A])] .
	\end{align*}
	
	We will now apply this equation twice, first to the tours $T^v$ and then to an optimum solution $T^*$  to the original instance. 
	This way we can evaluate how good the solutions $T^v$ are for the original instance:
	\begin{align*}
		&\Exp_{A\sim p}[c(T^v[A])]\\[1mm]
		&\ =\ \sum_{U\subseteq V} \Prob_{A\sim p}[A=U]\cdot c(T^v[U]) \\
		&\ \le \ \sum_{U\subseteq V} \Prob_{A\sim p}[A=U]\cdot c(T^v[U\cup \{v\}])\\
		& \ = \ \Exp_{A\sim p^{v=d}}[c(T^v[A])] \\
		& \ \le \ \rho\cdot \Opt^{v=d} \\[1mm]
		&\ \le \ \rho\cdot \Exp_{A\sim p^{v=d}}[c(T^*[A])] \\[1mm]
		& \ = \ \rho\cdot \sum_{U\subseteq V} \Prob_{A\sim p}[A=U]\cdot c(T^* [U\cup\{v\}]) \\
		& \ \le \ \rho\cdot \sum_{U\subseteq V} \Prob_{A\sim p}[A=U]\cdot c(T^* [U]) 
		+ \rho\cdot \sum_{\substack{U\subseteq V\\ U\setminus\{v\}\neq \emptyset}} \Prob_{A\sim p}[A=U]\cdot \left(c(v^-_U,v)+c(v,v^+_U)\right), 
	\end{align*}
	where $v^-_U$ and $v^+_U$ are the first customers in $U\setminus\{v\}$ that we encounter when we traverse a fixed orientation of $T^*$ from $v$ 
	in backward and in forward direction, respectively. Given $u\neq w\in T^*$, we denote the set of customers (other than $u$ and $w$) that we visit when traversing $T^*$ in the given orientation from $u$ to $w$ by $T^*_{(u,w)}$.
	Taking the weighted average of these bounds, where the bound for $T^v$ is weighted with $\frac{p(v)}{p(V)}$,
	we conclude that the expected cost of the best such solution $T^v$ is at most
	\begin{align*}
\!\!\!\!\!\!\! 	\rho\cdot \sum_{U\subseteq V} \Prob_{A\sim p}[A=U]\cdot c(T^* [U]) 
		+ \sum_{v\in V} \frac{\rho\cdot  p(v) }{p(V)} \cdot \!\!
		\sum_{\substack{U\subseteq V\\ U\setminus\{v\}\neq \emptyset}} \Prob_{A\sim p}[A=U]\cdot \left(c(v^-_U,v)+c(v,v^+_U)\right). 
	\end{align*}
	
	Now for any ordered pair $(u,w)\in V^2$, the term $c(u,w)$ appears in the right-hand summand with coefficient
	\begin{align*}
		& \frac{\rho\cdot p(w)}{p(V)} \cdot \Prob_{A\sim p}[w^-_A=u] + \frac{\rho\cdot p(u)}{p(V)} \cdot \Prob_{A\sim p}[u^+_A=w] \\
		&  \ = \ \frac{2\rho}{p(V)}\cdot p(u)\cdot p(w)\cdot \!\prod_{v\in T^*_{(u,w)}}\! (1-p(v)) \\
		&  \ = \ \frac{2\rho}{p(V)}\cdot \Prob_{A\sim p} [u \in A,\, w=u^+_A] \\[1mm]
		&  \ \le \ \epsilon \cdot \Prob_{A\sim p} [u\in A,\, w=u^+_A]
	\end{align*}
	because $p(V)\ge \frac{2\rho}{\epsilon}$. Summing over all ordered pairs $(u,w)$ yields a bound of $ \epsilon \cdot \Exp_{A\sim p}[c(T^*[A])]$.
	Hence the expected cost of the best $T^v$ is at most
	\begin{align*}
		\left( \rho + \epsilon \right)\cdot \Exp_{A\sim p}[c(T^*[A])]
	\end{align*}
	as required. 
	
	It is easy to compute the exact expected cost of each $T^v$ deterministically in $O(|V|^3)$ time
	(and thus choose the best) via
	\begin{align*}
	\!	\Exp_{A\sim p}[c(T^v[A])] \ =  \sum_{\{u,w\}\in {V\choose 2}} p(u)\cdot p(w)\cdot 
		\left( \!\prod_{t\in T^v_{(u,w)}}\! (1-p(t)) + \!\prod_{t\in T^v_{(w,u)}}\! (1-p(t)) \right) \cdot c(u,w). 
	\end{align*}
\end{proof}

The proof of \cref{thm:can_assume_depot} follows from combining 
\cref{lemma_low_activity} and \cref{lemma:depot_when_highactivity}.

\section{Reducing to normalized instances (Proof of \cref{lemma:wlog_normalized})} 
\label{section:reduce_to_normalized}

In this section we prove \cref{lemma:wlog_normalized}, i.e., that upper bounds on the master route ratio and the approximation ratio of the sampling algorithm for normalized instances with low uniform activation probability yield the same bounds for all instances of the \textsc{a priori TSP} with a depot. The high-level idea of this reduction is rather simple: we replace each customer by many copies, each with the same very low activation probability, such that for each customer the probability that at least one of its copies is active roughly matches the probability that the original customer is active. However, as we will see, it requires quite some technical care to prove \cref{lemma:wlog_normalized} formally.

We first prove a few auxiliary statements that will be useful later.
\begin{proposition}\label{prop:natural_logarithm}
	Let $0<x < y<1$. Then 
\[
\frac{x}{y}-x \ < \ \frac{\ln(1-x)}{\ln(1-y)} \ < \ \frac{x}{y} .
\]
	\end{proposition}
\begin{proof}
For the first inequality, we calculate 
\[-\ln(1-x) \ = \ \int_{1-x}^1\frac{1}{t}\ dt \ > \ x\enskip\text{ and }\enskip-\ln(1-y) \ = \ \int_{1-y}^1\frac{1}{t}\ dt \ < \ \frac{y}{1-y}.\]

For the second inequality, we compute
\begin{align*}
\quad-\ln(1-y) \ &=\ \int_{1-y}^{1}\frac{1}{t}\ dt \\ &> \ \int_{1-y}^{1-x}\frac{1}{1-x}\ dt+\int_{1-x}^1 \frac{1}{t}\ dt \\
&= \ \frac{y-x}{1-x}+\int_{1-x}^1 \frac{1}{t}\ dt \\
&> \ \frac{y-x}{x}\cdot\int_{1-x}^1 \frac{1}{t}\ dt+\int_{1-x}^1 \frac{1}{t}\ dt \\
&= \ \frac{y}{x}\cdot\int_{1-x}^1 \frac{1}{t}\ dt\\
&= \ -\frac{y}{x}\cdot\ln(1-x) . \qedhere
\end{align*}
\end{proof}
\begin{lemma}
	Let $\sigma\in (0,1)$, $\lambda > 0$ and $p\in (0,1]$. Then there is $\epsilon_p\in (0,1)$ with the following property:
	For every $\epsilon \in (0,\epsilon_p]$ there is $k \in \mathbb{N}$ such that 
	\begin{enumerate}
		\item $1-(1-\epsilon)^k\leq p\leq \epsilon k$, \label{aux_prob_item_1}
		\item $(1-p)^\sigma\leq(1-\sigma\epsilon)^k$, \label{aux_prob_item_2}
		\item $1-(1-p)^\sigma\leq (1+\lambda)\cdot (1-(1-\sigma\epsilon)^k)$, \label{aux_prob_item_3}
		\item $(1-\epsilon)^k\leq (1+\lambda)\cdot (1-p)$ if $p<1$, and $(1-\epsilon)^k\leq \lambda$ if $p=1$. \label{aux_prob_item_4}
	\end{enumerate}\label{lemma:auxiliary_probabilities}
\end{lemma}
\begin{proof} For $\epsilon \in (0,1)$ choose 
	\[k \ \coloneqq \ \begin{cases}\left\lfloor \frac{\ln(1-p)}{\ln(1-\epsilon)}\right\rfloor  & \text{if } p <1\\
		&\\
		\left\lceil\max\left\{\frac{1}{\epsilon},\frac{\ln(\frac{\lambda}{1+\lambda})}{\ln(1-\sigma\epsilon)}\right\}\right\rceil & \text{if }p=1\end{cases}.\] 
	We first handle the easier case $p=1$, in which the inequalities hold for any fixed $\epsilon\in (0,1)$. Indeed, the first inequality in \ref{aux_prob_item_1} and \ref{aux_prob_item_2} follow directly from $p=1$. Moreover, the second inequality in \ref{aux_prob_item_1} is implied by our choice of $k$. Finally, the definition of $k$ yields
	\[(1-\epsilon)^k\leq (1-\sigma\epsilon)^k \leq \frac{\lambda}{1+\lambda}<\lambda,\] which implies \ref{aux_prob_item_3} and \ref{aux_prob_item_4}.
	
	Now, we deal with the more interesting case $p<1$.
Applying \cref{prop:natural_logarithm} with $x=\frac{p}{2}$ and $y=p$ yields 
\[-\ln(1-p) \ > \ -2\ln(1-\textstyle{\frac{p}{2}}) \ > \ 0. \] 
Thus, there is $\epsilon_1>0$ such that 
\[
-\ln(1-p) \ \ge \ (1+\epsilon_1)\cdot(-2)\cdot\ln(1-\textstyle{\frac{p}{2}}).
\] 
As $p>0$, we have $(1-p)^\sigma < 1$, so there is $\kappa >0$ with 
\begin{equation}
1-(1+\kappa)\cdot (1-p)^\sigma \ \geq \ (1+\lambda)^{-1}\cdot (1-(1-p)^\sigma).\label{eq:aux_0}
\end{equation} 
Pick $\epsilon_2>0$ such that 
\[(1-\sigma\epsilon_2)^{-1}\cdot (1-p)^{-\sigma\epsilon_2} \ \leq \ 1+\kappa.\] 
Finally, let  $\epsilon_p\coloneqq\min\{p\cdot\epsilon_1,\epsilon_2,\frac{p}{3},\frac{\lambda}{1+\lambda}\}$. 
Now let $0<\epsilon\leq \epsilon_p$. Then we have
\begin{equation}
\frac{\ln(1-p)}{\ln(1-\epsilon)} \ \geq \ \frac{p}{\epsilon}\cdot \left(1+\frac{\epsilon}{p}\right)\label{eq:aux_1}
\end{equation} 
because  
\[\left(1+\frac{\epsilon}{p}\right)\cdot \frac{-\ln(1-\epsilon)}{\epsilon}
\ \leq \ (1+\varepsilon_1) \cdot \frac{-\ln(1-\epsilon)}{\epsilon}
\ < \ (1+\epsilon_1)\cdot \frac{-\ln(1-\frac{p}{2})}{\frac{p}{2}}\leq \frac{-\ln(1-p)}{p},
\] 
where we used $\varepsilon_p \le p \cdot \varepsilon_1$ in the first inequality, \cref{prop:natural_logarithm} with $x = \epsilon \le \frac{p}{3}< \frac{p}{2}=y$ in the second inequality, and the definition of $\epsilon_1$ in the third inequality.
By definition of $\epsilon_p$ and $\epsilon_2$, we also have
\begin{equation}
(1-\sigma\epsilon)^{-1}\cdot (1-p)^{-\sigma\epsilon} \ \leq \ 1+\kappa\label{eq:aux_2}.
\end{equation}

To prove \ref{aux_prob_item_1}, we compute 
\begin{align*}1-(1-\epsilon)^k&
\ \leq \ 1-(1-\epsilon)^{ \frac{\ln(1-p)}{\ln(1-\epsilon)}} \\
&\ = \ p\\ 
&\ = \ \epsilon\cdot\left(\frac{p}{\epsilon}\cdot\left(1+\frac{\epsilon}{p}\right)-1\right) \\
&\ \stackrel{\eqref{eq:aux_1}}{\leq} \ \epsilon\cdot \left(\frac{\ln(1-p)}{\ln(1-\epsilon)}-1\right) \\
&\ \leq \ \epsilon\cdot k.
\end{align*}

To prove \ref{aux_prob_item_2}, we calculate
\[(1-\sigma\epsilon)^k \ \geq \ (1-\sigma\epsilon)^{\frac{\ln(1-p)}{\ln(1-\epsilon)}} \ = \ (1-p)^{\frac{\ln(1-\sigma\epsilon)}{\ln(1-\epsilon)}}
\ \geq \ (1-p)^{\frac{\sigma\epsilon}{\epsilon}}=(1-p)^\sigma.\]
Here, the last inequality follows since $1-p\in [0,1)$ and $\frac{\ln(1-\sigma\epsilon)}{\ln(1-\epsilon)} < \frac{\sigma\cdot\epsilon}{\epsilon}$ by 
\cref{prop:natural_logarithm}. 

To prove \ref{aux_prob_item_3}, we compute
\begin{align*}
(1-\sigma \epsilon)^k&\ \leq \ (1-\sigma\epsilon)^{-1}\cdot (1-\sigma\epsilon)^{\frac{\ln(1-p)}{\ln(1-\epsilon)}} \\
&\ = \ (1-\sigma\epsilon)^{-1}\cdot(1-p)^\frac{\ln(1-\sigma\epsilon)}{\ln(1-\epsilon)}\\
&\ \leq \ (1-\sigma\epsilon)^{-1}\cdot (1-p)^{\sigma\cdot(1-\epsilon)}\\
&\ \stackrel{\eqref{eq:aux_2}}{\leq } \ (1+\kappa)\cdot (1-p)^\sigma,
\end{align*}
where the second inequality follows from \cref{prop:natural_logarithm} with $x=\sigma\epsilon$ and $y=\epsilon$.
Hence, \[1-(1-\sigma \epsilon)^k \ \geq \ 1-(1+\kappa)\cdot (1-p)^\sigma
\ \stackrel{\eqref{eq:aux_0}}{\geq} \ (1+\lambda)^{-1}\cdot (1-(1-p)^\sigma).\]

Finally, to prove \ref{aux_prob_item_4}, we calculate
\[ \!\! (1+\lambda)^{-1}\cdot(1-\epsilon)^k \ = \ \left(1-\frac{\lambda}{1+\lambda}\right)\cdot (1-\epsilon)^k 
\ \leq \ (1-\epsilon)^{k+1} \ \leq \ (1-\epsilon)^{\frac{\ln(1-p)}{\ln(1-\epsilon)}} \ = \ 1-p.\]
\end{proof}

\begin{lemma} \label{lemma:smaller_probs_decrease_opt}
Let $(V,c,p)$ be an instance of \textsc{a priori TSP} with depot $d$ and let $T^*$ be an optimum a priori tour for this instance. 
Let $p': V\to (0,1]$ such that $p'(v)\leq p(v)$ for all $v\in V$ and $p'(d)=p(d)=1$. 
Then 
\[\Opt(V,c,p') \ \le \ \Exp_{A\sim p'}[c(T^*[A])] \ \leq \ \Exp_{A\sim p}[c(T^*[A])] \ = \ \Opt(V,c,p).\]
\end{lemma}

\begin{proof}
It suffices to prove the lemma for the special case where $p$ and $p'$ only differ in exactly one customer 
because the general case follows by induction. 
Thus, let $(V,c,p)$ be an instance of \textsc{a priori TSP} with depot $d$, 
let $w\in V\setminus \{d\}$ and let $p':V\to(0,1]$ such that $p'(w)\leq p(w)$ and $p'(v)=p(v)$ for all $v\in V\setminus\{w\}$. Pick an optimum a priori tour $T^*$ for $(V,c,p)$. Then $T^*$ is also feasible for $(V,c,p')$. We compute
\begin{align*}
&\Opt(V,c,p') \ \le \ \Exp_{A\sim p'}[c(T^*[A])]\\
&\ = \ \sum_{S\subseteq V\setminus\{w\}} \Prob_{A\sim p'}[A\setminus\{w\}=S]\cdot \left( p'(w)\cdot c(T^*[S\cup\{w\}])+(1-p'(w))\cdot c(T^*[S]) \big.\right)\\
&\ = \ \sum_{S\subseteq V\setminus\{w\}} \Prob_{A\sim p}[A\setminus\{w\}=S]\cdot \left( p'(w)\cdot c(T^*[S\cup\{w\}])+(1-p'(w))\cdot c(T^*[S]) \big.\right)\\
&\ \leq \ \sum_{S\subseteq V\setminus\{w\}} \Prob_{A\sim p}[A\setminus\{w\}=S]\cdot \left( p(w)\cdot c(T^*[S\cup\{w\}])+(1-p(w))\cdot c(T^*[S]) \big.\right)\\
&\ = \ \Exp_{A\sim p}[c(T^*[A])] \ = \ \Opt(V,c,p).
\end{align*}
The last inequality follows from $p'(w)\leq p(w)$ and $c(T^*[S])\leq c(T^*[S\cup\{w\}])$.
\end{proof}

\begin{lemma}\label{lemma:MR_with_depot}
Let $(V,c,p)$ be an instance of \textsc{a priori TSP} with depot $d$ and let $d\in S\subseteq V$.
Then
\[\MR(S) \ = \ \left(1-\prod_{v\in V\setminus \{d\}}(1-p(v))\right)\cdot\OptTSP(S,c)+2\cdot \sum_{v\in V\setminus\{d\}} p(v)\cdot c(v,S) .\] 
\end{lemma}
\begin{proof}
	By definition, we have
\begin{align*}
\MR(S)&\ = \ \Exp_{A\sim p} \left[ \mathds{1}_{|A|\ge 2} \cdot \left( \OptTSP(S,c)+2\cdot\sum_{v\in A}c(v,S) \right) \right]\\
&\ = \ \Prob_{A\sim p}[|A|\geq 2]\cdot\OptTSP(S,c)+2\cdot\Exp_{A\sim p}\left[\mathds{1}_{|A|\ge 2} \cdot\sum_{v\in A}c(v,S)\right].\end{align*}
As the depot is always active, we have $|A|<2$ if and only if every customer in $V\setminus\{d\}$ is inactive, which happens with probability 
$\prod_{v\in V\setminus \{d\}}(1-p(v))$. 
Hence, 
\[\Prob_{A\sim p}[|A|\geq 2] \ = \ 1-\prod_{v\in V\setminus \{d\}}(1-p(v)) .\]
We further observe that 
\[\Exp_{A\sim p}\left[\mathds{1}_{|A|\ge 2} \cdot\sum_{v\in A}c(v,S)\right] 
\ = \ \Exp_{A\sim p}\left[\sum_{v\in A\setminus\{d\}}c(v,S)\right]=\sum_{v\in V\setminus\{d\}}p(v)\cdot c(v,S),\]
where the first equality follows from the facts that $\sum_{v\in A}c(v,S)=\sum_{v\in A\setminus\{d\}}c(v,S)$ since $d\in S$ and that this sum can only be nonzero if $|A|\geq 2$ because $d$ is always active.
\end{proof}

Now we are ready to prove \cref{lemma:wlog_normalized}.

\begin{proof}[Proof of \cref{lemma:wlog_normalized}]
Let $(V,c,p)$ be an instance of \textsc{a priori TSP} with depot $d$. 
We show that for every $\delta >0$, there exist $i\in\mathbb{N}$ and an $\epsilon_i$-normalized instance $(V',c',p')$ with depot $d$ such that
\begin{enumerate}
	\item\label{prop:item:opt} $\Opt(V',c',p') \le \Opt(V,c,p)$; 
	\item\label{prop:item:mrr} $(1+\delta)\cdot\min\limits_{S' \subseteq V' : d \in S'} \MR(S') \ge \min\limits_{S \subseteq V : d \in S} \MR(S)$;
	\item\label{prop:item:sampling} Let $f(q)\coloneqq 1-(1-q)^\sigma$ and $f'(q)\coloneqq \sigma\cdot q$ for all $q\in [0,1)$, and $f(1)\coloneqq f'(1) \coloneqq 1$. 
	Then
	\begin{align*}
	 (1+\delta) \,\cdot\, &\Exp_{S\sim f'\circ p'}\left[\alpha\cdot\OptTSP(S,c')+2\cdot\sum_{v\in V'}p(v)\cdot c'(v,S)\right]\\ 
	\ \ge \ &\Exp_{S\sim f\circ p}\left[\alpha\cdot\OptTSP(S,c)+2\cdot\sum_{v\in V}p(v)\cdot c(v,S)\right].
	\end{align*}
\end{enumerate}
Note that this immediately gives \cref{lemma:wlog_normalized}.
Let $n \coloneqq |V|$ and pick $\lambda>0$ such that $1-\lambda \geq (1+\delta)^{-1}$, $(1+\lambda)^{n-1}\leq 1+\delta$ and 
\begin{equation} \label{eq:choice_lambda}
(1+\delta)\cdot\left(1-(1+\lambda)^{n-1}\cdot\prod_{v\in V\setminus\{d\}} (1-p(v))\right) \ \geq \ 1-\prod_{v\in V\setminus\{d\}}(1-p(v)).
\end{equation} 
Observe that this is possible because $\prod_{v\in V\setminus\{d\}} (1-p(v))<1$ since $p(v)>0$ for all $v\in V$, 
meaning that \eqref{eq:choice_lambda} is a strict inequality for $\lambda = 0$.
Apply \cref{lemma:auxiliary_probabilities} to obtain constants $\epsilon_{p(v)}>0$ for all $v\in V$. 
Pick $i$ such that $\epsilon_i\leq \min_{v\in V}\epsilon_{p(v)}$. 
 Let $(V',c',p')$ result from $(V,c,p)$ by keeping the depot $d$ with $p(d)=1$ unchanged,
and replacing each other $v\in V$ by $k_v$ copies of $v$, each with activation probability $\epsilon_i$, where $k_v$ is chosen as the number $k$ in \cref{lemma:auxiliary_probabilities} for $p=p(v)$ and $\epsilon = \epsilon_i$. 
Denote the projection of $V'$ onto $V$ by $\pi$. Then 
\begin{equation} \label{eq:projection}
\begin{aligned}
\OptTSP(S',c') &\ = \ \OptTSP(\pi(S'),c) \qquad \text{ and } \\
c'(v',S') &\ = \ c(\pi(v'),\pi(S')) \hspace*{11mm} \text{ for all } S'\subseteq V', v'\in V'.
\end{aligned}
\end{equation}

First, we show \ref{prop:item:opt}. Let $T^*$ be an optimum a priori tour for $(V,c,p)$ and let $T'$ arise from $T^*$ by replacing each $v\in V\setminus\{d\}$ by its $k_v$ copies. 
Then $T'$ is a feasible solution for $(V',c',p')$. Now, sample $S' \subseteq V'$ according to the activation probabilities $p'$. We always have $c'(T'[S'])=c(T^*[\pi(S')])$.
Define $\bar{p}(d) \coloneqq 1$ and $\bar{p}(v) \coloneqq 1-(1-\epsilon_i)^{k_v}$ for each $v \in V \setminus \{d\}$.
Note that $\bar{p}(v) \leq p(v)$ by \cref{lemma:auxiliary_probabilities}~\ref{aux_prob_item_1}. 
Moreover, for $v\in V$, the probability that $v\in\pi(S')$ equals
$\bar{p}(v)$.
Hence, the expected cost of $T'$ equals the expected cost of $T^*$ for the instance $(V,c,\bar{p})$.
Thus, \cref{lemma:smaller_probs_decrease_opt} allows us to conclude that $\Opt(V',c',p')\leq \Opt(V,c,p)$.

Next we show \ref{prop:item:mrr}.
Let $S'\subseteq V'$ with $d \in S'$ and let $S\coloneqq\pi(S')$.
Note that $\OptTSP(S,c)=\OptTSP(S',c')$.
We claim that
\begin{equation}
(1+\delta)^{-1}\cdot\left(1-\prod_{v\in V\setminus\{d\}} (1-p(v))\right)\ \leq \ 1-\prod_{v'\in V'\setminus\{d\}} (1-p'(v')). \label{eq:prob_at_least_2}
\end{equation} 
Indeed, if there is $w\in V\setminus\{d\}$ with $p(w)=1$, then \cref{lemma:auxiliary_probabilities}~\ref{aux_prob_item_4} tells us that
\begin{align*}
1 \ - \prod_{v'\in V'\setminus\{d\}} (1-p'(v')) &\ \geq \ 1-(1-\epsilon_i)^{k_w} \ \geq \ 1-\lambda \ \geq \ (1+\delta)^{-1} \\
&\ = \ (1+\delta)^{-1}\cdot\left(1-\prod_{v\in V\setminus\{d\}} (1-p(v))\right).
\end{align*}
On the other hand, if $p(v)<1$ for all $v\in V\setminus\{d\}$, then  \cref{lemma:auxiliary_probabilities}~\ref{aux_prob_item_4} and \eqref{eq:choice_lambda} tell us that
\begin{align*}
&(1+\delta)^{-1}\cdot\left(1-\prod_{v\in V\setminus\{d\}} (1-p(v))\right)\\
&\ \leq \ 1-\prod_{v\in V\setminus\{d\}} (1+\lambda)\cdot(1-p(v))\\
&\ \leq \ 1-\prod_{v\in V\setminus \{d\}} (1-\epsilon_i)^{k_v}\\
&\ = \ 1-\prod_{v'\in V'\setminus\{d\}} (1-p'(v')).
\end{align*}
By \cref{lemma:MR_with_depot} and \cref{lemma:auxiliary_probabilities}~\ref{aux_prob_item_1}, we obtain
\begin{align*}
\!\!\!\!\!\! \MR(S) 
&\ = \ \left(1- \!\!\prod_{v\in V\setminus \{d\}} \!\! (1-p(v))\right)\! \cdot\OptTSP(S,c)+2\cdot \!\!\!\!\sum_{v\in V\setminus\{d\}} \!\! p(v)\cdot c(v,S)\\
&\ \stackrel{\eqref{eq:prob_at_least_2}}{\leq}  (1+\delta)\cdot\left(1- \!\!\prod_{v'\in V'\setminus \{d\}} \!\! (1-p'(v'))\right)\! \cdot\OptTSP(S',c')+2\cdot \!\!\!\!\sum_{v\in V\setminus\{d\}} \!\! p(v)\cdot c(v,S)\\
&\ \leq \ (1+\delta)\cdot\left(1- \!\!\prod_{v'\in V'\setminus \{d\}} \!\! (1-p'(v'))\right)\! \cdot\OptTSP(S',c')+2\cdot \!\!\!\! \sum_{v\in V\setminus\{d\}} \!\!k_v\cdot\epsilon_i\cdot c(v,S)\\
&\ = \ (1+\delta)\cdot\left(1- \!\!\prod_{v'\in V'\setminus \{d\}} \!\! (1-p'(v'))\right)\! \cdot\OptTSP(S',c')+2\cdot \!\!\!\! \sum_{v'\in V'\setminus\{d\}} \!\! p'(v')\cdot c'(v',S')\\[1mm]
&\ = \ (1+\delta)\cdot \MR(S').
\end{align*}
Thus, $(1+\delta)\cdot\MR(S')\geq \MR(S)$.

Finally, we bound the sampling costs as in \ref{prop:item:sampling}. For every $U\subseteq V$ with $d\in U$, we have
\begin{align*}
\Prob_{S\sim f\circ p}[S=U] &\ = \ \prod_{v\in U\setminus\{d\}}(1-(1-p(v))^\sigma)\cdot\prod_{v\in V\setminus U}(1-p(v))^\sigma\\
&\ \le \ \prod_{v\in U\setminus\{d\}}(1+\lambda)\cdot (1-(1-\sigma\cdot\epsilon_i)^{k_v})\cdot\prod_{v\in V\setminus U}(1-\sigma\cdot\epsilon_i)^{k_v}\\
&\ \leq \ (1+\lambda)^{n-1}\cdot \Prob_{S'\sim f'\circ p'}[\pi(S')=U]\\[1mm]
&\ \leq \ (1+\delta)\cdot \Prob_{S'\sim f'\circ p'}[\pi(S')=U],
\end{align*}
where we used \cref{lemma:auxiliary_probabilities}~\ref{aux_prob_item_2} and \ref{aux_prob_item_3} in the first inequality.
We use this to compare the expected costs of the master tours:
\begin{align*}
\alpha\cdot \Exp_{S\sim f\circ p}[\OptTSP(S,c)]
&\ = \ \alpha\cdot \sum_{U\subseteq V}\Prob_{S\sim f\circ p}[S=U]\cdot \OptTSP(U,c)\\
&\ \leq \ (1+\delta)\cdot\alpha\cdot \sum_{U\subseteq V}\Prob_{S'\sim f'\circ p'}[\pi(S')=U]\cdot \OptTSP(U,c)\\
&\stackrel{\eqref{eq:projection}}{=}(1+\delta)\cdot\alpha\cdot \sum_{U'\subseteq V'}\Prob_{S'\sim f'\circ p'}[S'=U']\cdot \OptTSP(U',c')\\
&\ = \ (1+\delta) \cdot \alpha\cdot \Exp_{S'\sim f'\circ p'}[\OptTSP(S',c')].
\end{align*}
For the connection costs, we compute a bound of
\begin{align*}
&\Exp_{S\sim f\circ p}\left[\sum_{v\in V} 2\cdot p(v)\cdot c(v,S)\right] \\
&\ = \ \sum_{U\subseteq V}\Prob_{S\sim f\circ p}[S=U]\cdot \sum_{v\in V} 2\cdot p(v)\cdot c(v,U)\\
&\ \leq \ (1+\delta)\cdot\sum_{ U\subseteq V} \Prob_{S'\sim f'\circ p'}[\pi(S')=U]\cdot\sum_{v\in V} 2\cdot p(v)\cdot c(v,U)\\
&\ \le \ (1+\delta)\cdot\sum_{ U\subseteq V} \Prob_{S'\sim f'\circ p'}[\pi(S')=U]\cdot\sum_{v\in V} 2\cdot \epsilon_i \cdot k_v \cdot c(v,U)\\
&\stackrel{\eqref{eq:projection}}{=}  (1+\delta)\cdot\sum_{ U'\subseteq V'} \Prob_{S'\sim f'\circ p'}[S'=U']\cdot \sum_{v\in V'} 2\cdot p'(v)\cdot c'(v,U') \\
&\ = \ (1+\delta)\cdot \Exp_{S'\sim f'\circ p'}\left[\sum_{v\in V'} 2\cdot p'(v)\cdot c'(v,S')\right] 
\end{align*}
where we used \cref{lemma:auxiliary_probabilities}~\ref{aux_prob_item_1} in the second inequality.
This proves \ref{prop:item:sampling}.
\end{proof}

\section{Analysis of the deterministic algorithm via the master route ratio}
\label{section:deterministic_via_masterrouteratio}

Our new upper bound on the master route ratio implies a better deterministic approximation algorithm as the following theorem shows.
This is essentially due to van Zuylen in \cite{Zuy11}, although she did not formally define the master route ratio.

\begin{theorem}  \label{thm:masterrouteratio_implies_deterministic}
	Let $\rho$ denote the master route ratio for \textsc{a priori TSP} instances with depot.
	Suppose we have an algorithm for (metric) TSP that always computes a tour of cost at most $\alpha$ times the value of the subtour elimination LP.
	Then there is a deterministic $(2+\alpha\rho)$-approximation algorithm for \textsc{a priori TSP} instances with depot.
\end{theorem}

By \cref{thm:mrr} we have $\rho<2.6$. Together with \cref{thm:can_assume_depot}, 
plugging in $\alpha = 1.5$ \cite{Wol80} or $\alpha = 1.5 - 10^{-36}$ \cite{KarKO23}, 
yields \cref{cor:approximation_ratio_deterministic}. The proof of \cref{thm:masterrouteratio_implies_deterministic} follows van Zuylen \cite{Zuy11}:

\begin{proof}
	In order to derandomize the approximation algorithm for \textsc{a priori TSP} sketched in \cref{sec:our_results}, 
	it suffices to determine the nonempty set $S$ on which we build the master tour in a deterministic way.
	In light of \cref{lemma:MR_with_depot}, our goal is to find a nonempty set $S$ and a TSP tour $T$ for $S$ minimizing
	\[ \left(1-\prod_{v\in V\setminus \{d\}} (1-p(v))\right) \cdot c(T) + \sum_{v\in V}2 p(v) c(v,S).\]
	Using the method of conditional expectation, we decide for each customer one by one whether it should be part of $S$. 
	We maintain a set $P$ of customers chosen to be in $S$ and a set $\overline{P}$ of customers that will not be part of $S$. 
	Initially, $P=\{d\}$ and $\overline P=\emptyset$. 
	Considering a customer $v\notin P\cup\overline{P}$, we compute a pessimistic estimator for the conditional expectation value of 
	\[ \left(1-\prod_{v\in V\setminus \{d\}} (1-p(v))\right) \cdot c(T) + \sum_{v\in V}2 p(v) c(v,S) \] 
	when drawing $S$ according to the activation probabilities 
	-- once under the condition that $P\cup\{v\}\subseteq S$ and $\overline{P}\cap S = \emptyset$ 
	and once under the condition that $P\subseteq S$ and $(\overline{P}\cup \{v\})\cap S = \emptyset$. 
	We add $v$ to $P$ or $\overline{P}$ depending on which estimator for the conditional expectation is smaller.
	
	The pessimistic estimator has two components. 
	First, the expected cost for connecting the active customers to the master tour on $S$, where $P\subseteq S\subseteq V\setminus\overline{P}$ 
	and every $v\in V\setminus(P\cup\overline{P})$ is independently included into $S$ with probability $p(v)$, is
	\begin{equation} \label{eq:anke_connectioncost}
		\sum_{v \in V} 2 p(v) \cdot \Exp_{S\sim p} \left[ c(v,S) \mid P\subseteq S\subseteq V\setminus\overline{P} \right].
	\end{equation}
	It is not hard to see that this can be computed exactly in polynomial time for any $P,\overline{P}$. 
	To bound the conditional expectation of the cost of the master tour, let $E={V\choose 2}$ and
	consider the following linear program, where
	$Q\coloneqq\left(1-\prod_{v\in V\setminus \{d\}}(1-p(v))\right)$
	is the probability that the depot is not the only active customer:
	\begin{align}
		\min \sum_{e\in E}\left( Q \cdot c(e)b_e + \sum_{v\in V\setminus\{d\}}p(v)c(e)r_e^v\right) \tag{Master-Route-Solution-LP} \label{LP:MRLP}\\
		\text{subject to }\sum_{e\in \delta(U)}(b_e + r_e^v)  & \geq 2 & \text{ for } v\in  U\subseteq V\setminus\{d\} \notag \\
		b_e, r_e^v &\geq 0 &\text{ for } e\in E, v\in V\setminus\{d\}. \notag
	\end{align}
	The variables $b_e$ represent the edges of the master tour (think of them as edges that we \textbf{b}uy) 
	and the variables $r_e^v$ stand for the edges we only use to connect $v$ to the master tour if $v$ is active 
	(think of them as edges that we \textbf{r}ent for every customer separately). 
	By \cref{lemma:MR_with_depot},
	\eqref{LP:MRLP} is an LP relaxation of the problem of finding the best master route solution.
	From the definition of the master route ratio it follows that the cost of an optimum solution to \eqref{LP:MRLP} 
	is most $\rho\cdot\Opt$, where $\Opt$ again denotes the expected cost of an optimum a priori tour.
	
	The LP can be solved in polynomial time by standard techniques. 
	Given an optimum solution $\hat b,\hat r$ to \eqref{LP:MRLP}, we 
	set $Q_{P,\overline{P}}\coloneqq \Prob_{S\sim p} \left[ |S|\ge 2 \mid P\subseteq S\subseteq V\setminus\overline{P} \right]$
	and claim that
	\begin{equation} \label{eq:anke_masterroutebound}
		\alpha\cdot \sum_{e\in E} c(e) \left( Q_{P,\overline{P}} \cdot \hat b_e + \sum_{v\in P\setminus \{d\}} \hat r^v_e + \sum_{v\in V\setminus(P\cup\overline{P})}p(v)\hat r^v_e \right)
	\end{equation}
	is an upper bound on the conditional expected cost of the master tour. 
	To show this, note that \eqref{eq:anke_masterroutebound} is 
	\[\Exp_{S\sim p} \left[\left. \alpha\cdot \sum_{e\in E}c(e) y^S_e \ \right| \ P\subseteq S\subseteq V\setminus \overline{P} \right],\]
	where $y^{\{d\}}\coloneqq0$ and
	$y^S$ for $\{d\}\subsetneq S\subseteq V$ is defined by $y^S_e = \hat b_e + \sum_{v\in S\setminus\{d\}}\hat r_e^v$ for $e\in E$.
	For all $S$ with $|S|\ge 2$, we have that $y^S$ is a feasible solution to the subtour elimination LP
	\begin{align*}
		\min \sum_{e\in E}c(e)y_e \\ 
		\text{subject to } \sum_{e\in \delta(U)}y_e  &\geq 2   & \text{ for } U\subset V: S\setminus U\neq \emptyset, S\cap U\neq \emptyset \\
		y_e &\geq 0 &\text{ for } e\in E . 
	\end{align*}
	Hence for given $S$ we can find (in polynomial time) a TSP tour for $S$ that has cost at most $\alpha\cdot  \sum_{e\in E}c(e) y^S_e$,
	and \eqref{eq:anke_masterroutebound} is the conditional expectation of this upper bound.
	Therefore we can use the sum of \eqref{eq:anke_masterroutebound} and \eqref{eq:anke_connectioncost} as pessimistic estimator.
	
	For $P=\{d\}$ and $\overline{P}=\emptyset$, we note that $Q=Q_{\{d\},\emptyset}$.
	Moreover, in this case, \eqref{eq:anke_connectioncost} is at most $2\cdot\Opt$ (as in \cref{sec:overview}),
	and
	\eqref{eq:anke_masterroutebound} is $\alpha \sum_{e\in E}\left( Q\cdot c(e)\hat b_e + \sum_{v\in V \setminus \{d\}}p(v) c(e) \hat r_e^v\right)$,
	which is at most $\alpha\rho\cdot\Opt$ as we observed above.
	The conditional expectation never increases during the described procedure 
	as the current value of the conditional expectation is always a convex combination of the two possible next values. 
	Thus the described deterministic algorithm results in an $(\alpha\rho + 2)$-approximation.
\end{proof}

\section{Discussion}

We conjecture (but could not prove) that our lower bound examples are really worst-case examples, and that the values of our 
linear programs converge to these bounds. 

Another question is whether the master route ratio is $\frac{1}{1-e^{-1/2}}$ even for low-activity instances.
Currently we only know the upper bound of $3$ from \cite{ShmT08}, but know no example with
master route ratio larger than $\frac{1}{1-e^{-1/2}}$ (and this value is attained by our example only as the activity tends to infinity).
The analogous question applies to the sampling algorithm:
whether we need to consider the low-activity case separately is an open question.

Finally, we hope that our approach can also help for proving a better bound for related problems where similar random sampling techniques are used,
or for showing that known bounds are best possible.

\appendix
\FloatBarrier

\bibliographystyle{plainurl}
\bibliography{aptsp.bib}

\end{document}